\documentclass[ %
12pt,
a4paper,
ngerman,
english,
]{scrartcl} %

\usepackage{babel}              
\usepackage[utf8]{inputenc}   
\usepackage[T1]{fontenc}        
\usepackage[scaled]{helvet}     

\usepackage{amsmath}
\usepackage{amsfonts}
\usepackage{amssymb}
\usepackage{amsthm}    
\usepackage{graphicx}   
\usepackage{caption}
\usepackage{subcaption}
\usepackage{geometry} 
\usepackage[round,comma,sort]{natbib} 
\usepackage{color}
\usepackage{epstopdf}           
\usepackage{fancybox}           
\usepackage{tabu}
\usepackage{hhline}
\usepackage[draft]{hyperref}
\hypersetup{final}
\usepackage{cleveref}
\usepackage{makeidx}            
\usepackage{enumerate}          
\usepackage{doi}
\makeindex

\newif\ifPics %
\Picstrue 

\begin{document}
\title{Investigating the Pilot Point Ensemble Kalman Filter for
  geostatistical inversion and data assimilation} %

\author{Johannes Keller%
  \thanks{ %
    Corresponding Author: jkeller at eonerc.rwth-aachen dot de, %
    Institute for Applied Geophysics and Geothermal Energy, %
    E.ON Energy Research Center, %
    RWTH Aachen University, %
    Aachen, %
    Germany %
  }
  \footnotemark[2] %
  \footnotemark[3] %
  \and Harrie-Jan Hendricks Franssen%
  \thanks{ %
    Forschungszentrum J{\"u}lich GmbH, %
    Institute of Bio- and Geosciences, %
    IBG-3 (Agrosphere), %
    J{\"u}lich, %
    Germany %
  } %
  \thanks{ %
    Centre for High-Performance Scientific Computing in Terrestrial
    Systems (HPSC-TerrSys), %
    Geoverbund ABC/J, %
    J{\"u}lich, %
    Germany %
  } %
  \and Wolfgang Nowak%
  \thanks{ %
    Institute for Modelling Hydraulic and Environmental
    Systems (LS3)/SimTech, %
    University of Stuttgart, %
    Stuttgart, %
    Germany%
  } %
} %

\maketitle

Accepted for publication in Advances in Water Resources. Copyright
2021 Elsevier. Further reproduction or electronic distribution is not
permitted.
%

\begin{abstract}
  Parameter estimation has a high importance in the geosciences. %
  The ensemble Kalman filter (EnKF) allows parameter estimation for
  large, time-dependent systems. %
  For large systems, the EnKF is applied using small ensembles, which
  may lead to spurious correlations and, ultimately, to filter
  divergence. %
  We present a thorough evaluation of the pilot point ensemble Kalman
  filter (PP-EnKF), a variant of the ensemble Kalman filter for
  parameter estimation. %
  In this evaluation, we explicitly state the update equations of the
  PP-EnKF, discuss the differences of this update equation compared to
  the update equations of similar EnKF methods, and perform an
  extensive performance comparison.
  The performance of the PP-EnKF is tested and compared to the
  performance of seven other EnKF methods in two model setups, a
  tracer setup and a well setup. %
  In both setups, the PP-EnKF performs well, ranking better than the
  classical EnKF. %
  For the tracer setup, the PP-EnKF ranks third out of eight
  methods. %
  At the same time, the PP-EnKF yields estimates of the ensemble
  variance that are close to EnKF results from a very large-ensemble
  reference, suggesting that it is not affected by underestimation of
  the ensemble variance. %
  In a comparison of the ensemble variances, the PP-EnKF ranks first
  and third out of eight methods. %
  Additionally, for the well model and ensemble size 50, the PP-EnKF
  yields correlation structures significantly closer to a reference
  than the classical EnKF, an indication of the method's skill to
  suppress spurious correlations for small ensemble sizes. %
\end{abstract}

\section{Introduction}
\label{sec:introduction}

Predictions of groundwater flow, mass transport and heat transport are
strongly influenced by subsurface hydraulic properties like the
hydraulic conductivity used in the groundwater flow equation. %
Unfortunately, hydraulic conductivity is often strongly heterogeneous
in space with limited measurement information. %
Starting in the 1970's, many studies worked on estimating these
properties with the help of inverse algorithms. %
The 1980's saw a transition towards stochastic inverse methods
\citep{Kitanidis1983} and, since the 1990's, methods were formulated
for generating multiple equally likely solutions to the groundwater
inverse problem \citep{GomezHernandez1997}.  %
In the 2000's, the ensemble Kalman filter method
\citep[EnKF,][]{Evensen1994, Burgers1998} became popular. %
In its parameter estimation version, it also calculates equally likely
solutions to the groundwater inverse problem, but avoids the
formulation of numerical derivatives that were needed in many
inversion methods used until then \citep[e.g.,][]{Chen2006,
  Hendricks2008, Nowak2009}. %
An overview on the groundwater inverse literature including comparison
of methods can be found in \citet{Carrera2005}, \citet{Hendricks2009}
and \citet{Zhou2014}. %

The ensemble Kalman filter is a sequential method, which is suited for
large and non-linear numerical models. %
When the EnKF is used for parameter estimation, parameters are updated
at a sequence of observation times according to observation data and
covariance information from an ensemble of stochastic realizations. %
After assimilation of all measurement data, the ensemble is used to
approximate the full posterior probability density distribution. %
However, problems such as filter inbreeding and ultimately filter
divergence have been diagnosed for the EnKF for small ensemble sizes
\citep{Hamill2001}. %

For many EnKF methods, sampling errors of the covariances between
dynamic variables and parameters are responsible for filter
divergence. %
A number of EnKF methods have been proposed and applied to tackle such
shortcomings of the EnKF related to small ensemble sizes. %
Two examples are the local EnKF \citep{Hamill2001} and the hybrid EnKF
\citep{Hamill2000}. %
In the local EnKF (also called covariance localization), covariances
are restricted to the vicinity of observations. %
In the hybrid EnKF, the full covariance matrix is a weighted sum of
the ensemble covariance matrix, which is the covariance of the
classical Kalman filter update, and a fixed covariance matrix, which
is defined before the assimilation, according to prior knowledge and
modeled the same way as the initial covariance matrix. %
Additional EnKF methods include the damped EnKF
\citep{Hendricks2008}, %
and the iterative EnKF \citep{Sakov2012}. %
In the damped EnKF, the proposed EnKF update is multiplied by a
damping factor to reduce the update of the parameters. %
This approach reduces problems with filter inbreeding. %
In contrast, the iterative EnKF restarts the whole assimilation after
every EnKF update, thereby reducing numerical instabilities connected
to nonlinear model equations. %

Further developments include the application of an ensemble smoother
\citep{Evensen2000, Chen2011, Emerick2013, Cosme2012}, and, in
particular, an iterative version of this smoother
\citep{Bocquet2013}. %
\citet{Crestani2013} compare ensemble smoothers to the EnKF in a
tracer test assimilation, concluding that the EnKF outperforms
smoothers in this scenario. %
Additionally, the iterative EnKF has been tested for additive model
error \citep{Sakov2018}. %
A disadvantage of iterative approaches is the strong increase in
needed compute time. %
Also, adaptive covariance inflation of the EnKF has been extensively
tested recently \citep{Raanes2019}. %
Another line of research focuses on the model equations and reducing
computational effort by combining EnKF variants with methods such as
principal component analysis and reduced basis \citep{Kang2017,
  Pagani2017, Xiao2018}. %
However, these methods do not tend to reduce problems with spurious
correlations and filter inbreeding. %

In this work, we explicitly formulate and extensively test the pilot
point ensemble Kalman filter (PP-EnKF) as an alternative approach for
parameter estimation. %
The PP-EnKF was initially introduced, sketched and tested in a small
suite of four tests by \citet{Heidari2013}, using a 50-member ensemble
in a petroleum reservoir model. %
The result was that this method yielded larger RMSE than classical
EnKF, getting better as the number of pilot points approached the full
set of parameters. %
On the other hand, spatial variability was found to be better
preserved by this method compared to the classical EnKF. %
Similar results were obtained by \citet[Chapter 4]{Crestani2013_2}. %
In a second performance comparison of the same method,
\citet{Tavakoli2013} tested the PP-EnKF from \citet{Heidari2013}
against the EnKF, ensemble smoothers and null-space Monte Carlo
methods using a 200-member ensemble in a multiphase model. %
In this comparison, the PP-EnKF performed well, yielding the smallest
RMSE for the estimated logarithmic permeability field. %
In \citep{Tavakoli2013}, the spread of the estimated EnKF-PP field was
actually smaller than for the other ensemble methods, a result that is
contradictory to the findings in \citep{Heidari2013}. %
This discrepancy makes it interesting to investigate this promising
method further, give an explicit statement of the mathematical
formulae and an extensive investigation of the performance of our
PP-EnKF. %

The specific contribution of this paper is twofold. %
First, the statement of the explicit mathematical formulae of the
PP-EnKF and its non-ensemble version, and second, an extensive testing
of the PP-EnKF in a large comparison. %
The original publication \citep{Heidari2013} contains no explicit
mathematical formulae, while our investigation starts at the classical
Kalman filter, shows the differences of the classical Kalman filter
and a pilot point Kalman filter and, finally, translates the
derivation to the corresponding ensemble methods obtaining the full
PP-EnKF filter equations and their ideal (linearized) estimation
variance. %
This derivation provides a rigorous mathematical and statistical
foundation to the method, and provides insights on its behavior in
various situations. %
Regarding the testing, our performance tests of the PP-EnKF are based
on a much larger test set compared to earlier tests in
\citep{Heidari2013} and \citep{Tavakoli2013}. %
Our statistically independent repetition of performance tests includes
RMSEs and overall standard deviations from 1,000 synthetic experiments
and, additionally, the comparison of full correlation fields from 10
synthetic experiments. %
The latter is to assess the ability of the PP-EnKF to suppress
spurious correlations. %

In Section \ref{sec:pp-enkf}, the PP-EnKF is introduced with a focus
on the differences between the classical EnKF and the PP-EnKF. %
In Section \ref{sec:design-syn-exp}, the two setups for synthetic
experiments and the performance evaluation measures are detailed. %
In Section \ref{sec:perf-comp}, we present results of the comparison
of the PP-EnKF to the other seven EnKF variants for the two parameter
estimation setups. %
Section \ref{sec:conclusion} contains a brief conclusion. %

\section{Pilot Point Ensemble Kalman Filter}
\label{sec:pp-enkf}

In this section, the ensemble equations of the PP-EnKF are introduced
by first defining the simpler equations of the pilot point Kalman
filter (PP-KF). %
The PP-KF is obtained by splitting the state vector of the Kalman
filter into dynamic variables, pilot point parameters and non-pilot
point parameters. %
The Kalman update is applied to dynamic variables and pilot point
parameters. %
The remaining parameter values are updated by a kriging interpolation
of the update. %
In the next step, the ensemble versions of the equations are
formulated, resulting in the PP-EnKF. %
We show how the update of the PP-EnKF, which is a mixture of an update
and an interpolation of this update, differs from the update in the
classical EnKF. %
Additionally, we compare the formulation of the PP-EnKF to the
formulations of the local EnKF and the hybrid EnKF in order to show
that, while aiming at the same goal of suppressing spurious
correlations, the PP-EnKF provides a useful alternative. %

\subsection{Kalman Filter and Ensemble Kalman Filter}
\label{sec:classical-enkf}

Before the PP-EnKF is presented, we quickly recall the equations for
the Kalman filter \citep{Kalman1960} and the ensemble Kalman filter
\citep{Evensen1994}. %
One advantage of the Kalman filter is that no adjoint equations are
needed in its formulation. %
The PP-EnKF differs from the EnKF only in the update equation. %
A full account of the EnKF can be found in \citet{Evensen2003} and
also in our earlier work, where we provide more detail on EnKF
variants and comparison methods \citep{Keller2018}. %

The unperturbed Kalman filter forward equation for the mean of the
state vector is given by %
\begin{equation}
  \label{eq:kalman-forward}
  \mathbf{x}^{f} = \mathbf{M}
  \mathbf{x}^{b},
\end{equation}
where the vector $\mathbf{x}^{b} \in R^{n_{s}}$ is either the initial
mean state vector or, in later time steps, the assimilated mean state
vector from the previous assimilation step. %
The forward computation is represented by the matrix
$\mathbf{M} \in R^{n_{s}\times n_{s}}$. %
We define $n_{s}$ as the size of the state vector. %
The state vector consists of both dynamic variables and static
parameters. %
$\mathbf{x}^{f} \in R^{n_{s}}$ is the state vector of predictions
computed by the forward simulation. %
In the Kalman filter, a second forward equation for the covariance
matrix is given by
  \begin{equation}
    \label{eq:kalman-forward-p}
    \mathbf{P} = \mathbf{M}
    \mathbf{P}^{b}\mathbf{M}^{T},
  \end{equation}
where $\mathbf{P}^{b} \in R^{n_{s}\times n_{s}}$ is either the initial
covariance matrix or the assimilated covariance matrix from the
previous assimilation step. %
$\mathbf{P} \in R^{n_{s}\times n_{s}}$ is the covariance matrix of the
predicted state vector. %
The predictions $\mathbf{x}^{f}$ and $\mathbf{P}$ serve as input to
the update equation of the Kalman filter. %

The Kalman filter update equation for the mean of the state vector is
given by %
  \begin{equation}
    \label{eq:kalman-update}
    \mathbf{x}^{a} - \mathbf{x}^{f}
    =
    \mathbf{P}\mathbf{H}^{T}
    \left(
      \mathbf{H}\mathbf{P}\mathbf{H}^{T}+\mathbf{R}
    \right)^{-1}
    \left(
      \mathbf{d}-\mathbf{H}\mathbf{x}^{f}
    \right)\, .
  \end{equation}
On the left-hand side of this equation, $\mathbf{x}^{a} \in R^{n_{s}}$
is the state vector after the Kalman filter update. %
$\mathbf{x}^{f} \in R^{n_{s}}$ is the state vector of predictions
computed by the forward simulation. %
On the right-hand side of the equation, $\mathbf{d} \in R^{n_{m}}$ is
the vector of measurements, where $n_{m}$ is the number of
measurements. %
$\mathbf{R} \in R^{n_{m}\times n_{m}}$ is the measurement error matrix
and $\mathbf{H} \in R^{n_{m}\times n_{s}}$ is the measurement
operator. %
The measurement operator maps the state vector to the measurements. %

The update equation for the covariance matrix
$\mathbf{P} \in R^{n_{s}\times n_{s}}$  %
  \begin{equation}
    \label{eq:kalman-cov-update}
    \mathbf{P}^{a} - \mathbf{P}
    =
    - \mathbf{P}\mathbf{H}^{T}
    \left(
      \mathbf{H}\mathbf{P}\mathbf{H}^{T}+\mathbf{R}
    \right)^{-1}
    \mathbf{H}\mathbf{P}
  \end{equation}
completes the Kalman filter update. %
$\mathbf{P}^{a} \in R^{n_{s}\times n_{s}}$ is the updated covariance
matrix. %

We now turn to the EnKF \citep{Evensen2003}. %
The EnKF is helpful for large and nonlinear models, where the
computation of the covariance matrix $\mathbf{P}$ of the Kalman filter
becomes unfeasible and nonlinearities yield large deviations in the
linear forward equations of the Kalman filter. %
The $n_{e}$ forward equations of the EnKF are given by
  \begin{equation}
    \label{eq:enkf-forward}
    \mathbf{x}_{i}^{f} = M
    \left(
      \mathbf{x}_{i}^{b}
    \right) , \qquad i \in
    \left\{
      1, \cdots, n_{e}
    \right\},
  \end{equation}
where $n_{e}$ is the size of the ensemble of state vector
realizations. %
The state vector realizations $\mathbf{x}_{i}^{b}$ make up either the
initial state vector ensemble or, in later time steps, the assimilated
state vector ensemble from the previous assimilation step. %
The state vector realizations $\mathbf{x}_{i}^{f}$ make up the
predicted state vector ensemble. %
The forward operator $M$ can be nonlinear, such as the numerical
solution of the groundwater flow equation. %
This makes the EnKF well adapted to nonlinear forward operators. %
However, the update equation itself is still linear and treats the
possibly altered probability distributions as multi-Gaussian. %
All covariance matrices in the EnKF are computed from the ensemble of
state vector realizations. %
The predictions $\mathbf{x}_{i}^{f}$ serve as input to the update
equation of the EnKF. %

The EnKF update equation is given by %
  \begin{equation}
    \label{eq:enkf-update}
    \mathbf{x}_{i}^{a} - \mathbf{x}_{i}^{f}
    =
    \mathbf{P}_{e}\mathbf{H}^{T}
    \left(
      \mathbf{H}\mathbf{P}_{e}\mathbf{H}^{T} + \mathbf{R}
    \right)^{-1}
    \left(
      \mathbf{d}_{i}-\mathbf{H}\mathbf{x}_{i}^{f}
    \right), \qquad i \in
    \left\{
      1, \cdots, n_{e}
    \right\}.
  \end{equation}
This equation is very similar to the Kalman filter update of the mean
state vector in Equation (\ref{eq:kalman-update}), with two important
differences. %
First, Equation (\ref{eq:enkf-update}) is computed $n_{e}$ times for
the realizations $\mathbf{x}^{f}_{i}$ and $\mathbf{x}^{a}_{i}$ of the
state vector, and realizations $\mathbf{d}_{i}$ of the measurements. %
Second, the covariance matrix $\mathbf{P}_{e}$ is estimated from the
ensemble of state vector realizations. %
Therefore, an analogue of update equation (\ref{eq:kalman-cov-update})
for the covariance matrix is not needed for the EnKF, since the
covariance matrix is updated by computing it from the update
ensemble. %
In this work, we write all update equations in terms of the difference
between an estimated quantity and the same quantity computed from the
forward simulation. %
When the Kalman filter update or the EnKF update is mentioned, this
refers to either this difference. %
Expressing the update equation with differences on the left-hand side
is convenient for introducing the pilot point EnKF (PP-EnKF). %
In the PP-EnKF, the updates on the left-hand sides of Equations
(\ref{eq:kalman-update}), (\ref{eq:kalman-cov-update}), and
(\ref{eq:enkf-update}) are the input of a kriging interpolation
\citep{Deutsch1992}. %

\subsection{Pilot Point EnKF}
\label{sec:algorithm}

Retracing the update equations in the last section, we will now
introduce the update equations of the pilot point Kalman filter
(PP-KF) and, subsequently, the update equations of the pilot point
ensemble Kalman filter (PP-EnKF). %
This way, the set of equations for the PP-EnKF can be rigorously
compared to those for the classical EnKF. %

The main idea of the PP-EnKF is to update in a first step only
parameters at a fixed subset of locations, called pilot points
\citep{RamaRao1995, GomezHernandez1997}. %
The positions of the pilot points are defined before the PP-EnKF
starts. %
In the rest of the model domain apart from the pilot points, parameter
values are initialized by random geostatistical simulation and are
then subject to an interpolated update. %
The interpolation is generated by ordinary kriging of the pilot point
updates and added to the geostatistically simulated fields. %
We use ordinary kriging, since we do not assume a background trend as
in universal kriging, or a correlation with other variables as in
cokriging. %
On the other hand, we expect the mean of the parameter update to vary
across the domain. %
The update of a certain parameter is defined as the difference between
the parameter value after the EnKF update step and before the EnKF
update step. %
While the updates apart from pilot points are interpolations of the
perturbations calculated at the pilot points, the full parameter
fields retain their initial spatial variability. %

\begin{figure}
  \centering \includegraphics[trim = 0 230 550 0, clip,
  width=0.8\columnwidth]{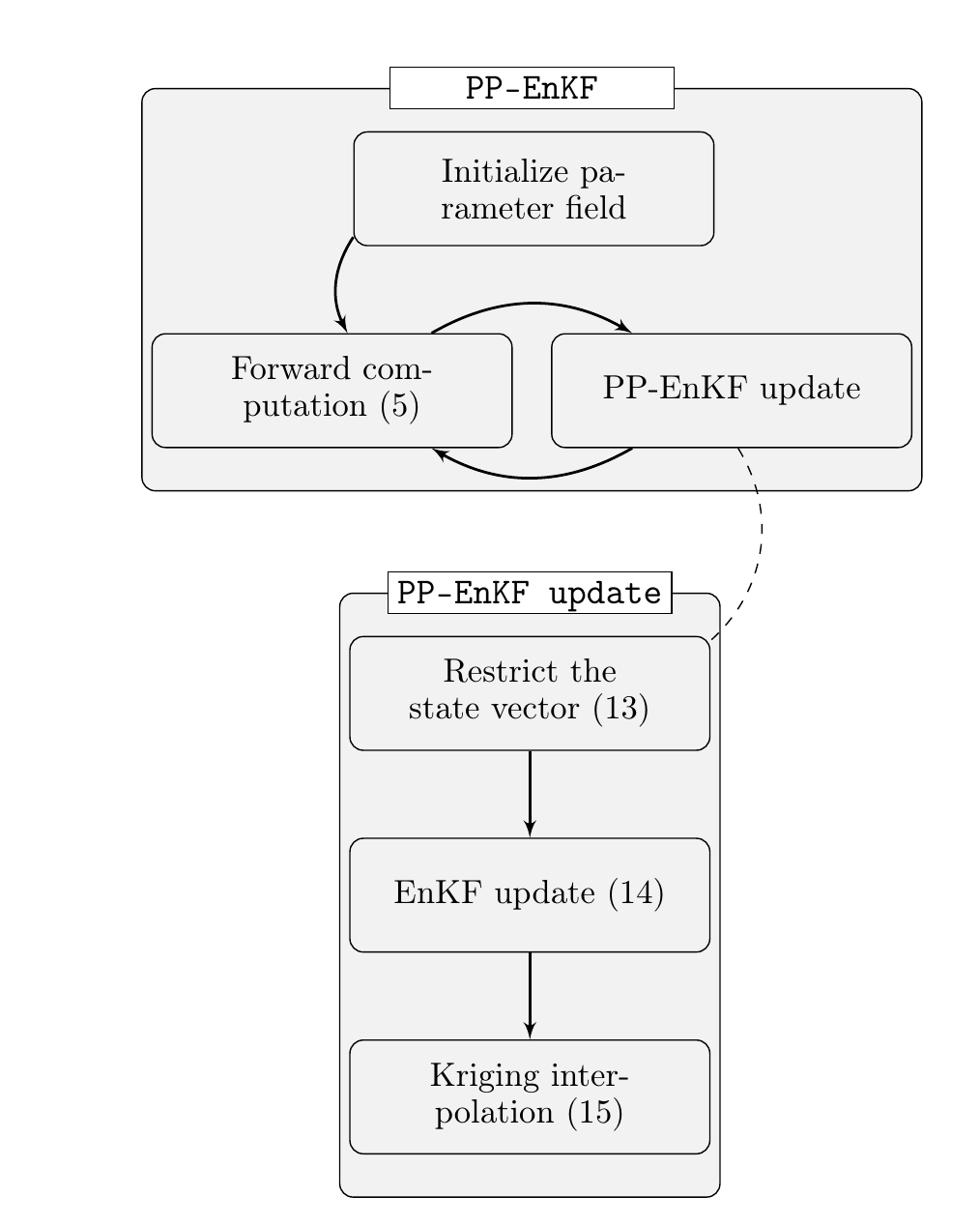}
  \caption{Diagram of the PP-EnKF workflow.  The numbers point to the
    equation of the corresponding step.}
  \label{fig:pp-enkf-diagram}
\end{figure}

In general, two covariance structures are important for the algorithm
of the PP-EnKF. %
First, there is the covariance structure used by the kriging
interpolation. %
This covariance structure is defined a priori and is reused for the
interpolation at each update step. %
Second, there is the covariance structure used for initial
geostatistical simulation. %
In the PP-EnKF, both covariance structures are equal and defined
according to prior knowledge. %
Overall, the partially fixed covariance structures should enable the
PP-EnKF to suppress spurious correlations, while still providing
ensemble-based updates that potentially affect the whole model
domain. %
Additionally, the geological realism of the ensemble is kept because
the initial randomization of the spatially heterogeneous field is
preserved, which should reflect the geology, and kriging only
interpolates the EnKF-based updates at the pilot points. %
We argue that these features make the PP-EnKF a useful alternative to
existing EnKF methods such as the hybrid EnKF and the local EnKF. %
Figure \ref{fig:pp-enkf-diagram} shows a diagram of the PP-EnKF. %

To define the update of the PP-KF, the state vector and its covariance
matrix are split into three parts, the pilot point parameters, the
non-pilot point parameters, and the dynamic variables. %
In order to introduce notation and make clear which part of the state
vector is left out in the PP-KF, we now rewrite the full state vector
and covariance matrix: %
  \begin{equation}
    \label{eq:ppkf-split}
    \mathbf{x}^{f} =
    \begin{pmatrix}
      \mathbf{x}^{f}_{p} \\ %
      \mathbf{x}^{f}_{r} \\ %
      \mathbf{x}^{f}_{d} %
    \end{pmatrix}
    \in R^{n_{s}} \qquad
    \mathbf{P} =
    \begin{pmatrix}
      \mathbf{P}_{pp} & \mathbf{P}_{pr} & \mathbf{P}_{pd} \\ %
      \mathbf{P}_{rp} & \mathbf{P}_{rr} & \mathbf{P}_{rd}\\ %
      \mathbf{P}_{dp} & \mathbf{P}_{dr} & \mathbf{P}_{dd}\\ %
    \end{pmatrix}
    \in R^{n_{s}\times n_{s}} \ .
  \end{equation}
Here, $\mathbf{x}^{f}_{p} \in R^{n_{p}}$ contains parameter values at
pilot point positions, and $n_{p}$ is the number of pilot point
parameters. %
$\mathbf{x}^{f}_{r} \in R^{n_{r}}$ denotes the parameters at non-pilot
point locations, where $n_{r}$ is the number of non-pilot point
parameters. %
Finally, $\mathbf{x}^{f}_{d} \in R^{n_{d}}$ contains the $n_{d}$
dynamic variables. %
The sum $n_{s} = n_{p}+n_{r}+n_{d}$ is the total number of entries in
the state vector. %
The covariance matrix $\mathbf{P}_{pp} \in R^{n_{p}\times n_{p}}$
contains the covariances between all pairs of parameters at pilot
points. %
$\mathbf{P}_{pr} \in R^{n_{p}\times n_{r}}$ and
$\mathbf{P}_{rp} \in R^{n_{r}\times n_{p}}$ contain covariances
between parameters at pilot points and parameters at non-pilot points.
$\mathbf{P}_{rr} \in R^{n_{r}\times n_{r}}$ contains covariances
between pairs of parameters at non-pilot points. %
$\mathbf{P}_{dp} \in R^{n_{d}\times n_{p}}$ and
$\mathbf{P}_{pd} \in R^{n_{p}\times n_{d}}$ contain covariances
between dynamic variables and parameters at pilot points. %
$\mathbf{P}_{dr} \in R^{n_{d}\times n_{r}}$ and
$\mathbf{P}_{rd} \in R^{n_{r}\times n_{d}}$ contain covariances
between dynamic variables and parameters at non-pilot points. %
Finally, $\mathbf{P}_{dd} \in R^{n_{d}\times n_{d}}$ contains
covariances between pairs of dynamic variables. %

The update of the PP-KF consists of two steps: updating the parameters
at pilot points and then interpolating the update. %
First, the pilot point update equations are discussed. %
The form of the mean vector update equation is identical to Equation
(\ref{eq:kalman-update}) of the Kalman filter, restricted to dynamic
variables and pilot-point parameters: %
  \begin{equation}
    \label{eq:ppkf-update-state}
    \begin{pmatrix}
      \mathbf{x}^{a}_{p} \\
      \mathbf{x}^{a}_{d}
    \end{pmatrix}
    -
    \begin{pmatrix}
      \mathbf{x}^{f}_{p} \\
      \mathbf{x}^{f}_{d}
    \end{pmatrix}
    =
    \begin{pmatrix}
      \mathbf{P}_{pp} & \mathbf{P}_{pd} \\
      \mathbf{P}_{dp} & \mathbf{P}_{dd}
    \end{pmatrix}
\mathbf{H}^{T} \left( \mathbf{P}_{yppy}
      + \mathbf{R} \right)^{-1}
    \left(
      \mathbf{d}-\mathbf{H}
      \begin{pmatrix}
        \mathbf{x}^{f}_{p} \\
        \mathbf{x}^{f}_{d}
      \end{pmatrix}
    \right) \ .
  \end{equation}
Explanations for the PP-KF versions of the covariance matrix among
simulated measurements $\mathbf{P}_{yppy} \in R^{n_{m}\times n_{m}}$
and the measurement operator
$\mathbf{H} \in R^{n_{m}\times (n_{p}+n_{d})}$ can be found in Section
\ref{sec:special-covariance}. %

The PP-KF update of the covariance matrix has the same form as
Equation (\ref{eq:kalman-cov-update}), but again restricted to dynamic
variables and parameters at pilot points: %
  \begin{equation}
    \label{eq:ppkf-update-cov}
    \begin{pmatrix}
      \mathbf{P}^{a}_{pp} & \mathbf{P}^{a}_{pd} \\
      \mathbf{P}^{a}_{dp} & \mathbf{P}^{a}_{dd}
    \end{pmatrix}
    -
        \begin{pmatrix}
      \mathbf{P}_{pp} & \mathbf{P}_{pd} \\
      \mathbf{P}_{dp} & \mathbf{P}_{dd}
    \end{pmatrix}
    = -
    \begin{pmatrix}
      \mathbf{P}_{pp} & \mathbf{P}_{pd} \\
      \mathbf{P}_{dp} & \mathbf{P}_{dd}
    \end{pmatrix}
    \mathbf{H}^{T} \left( \mathbf{P}_{yppy} + \mathbf{R}
    \right)^{-1} \mathbf{H}
    \begin{pmatrix}
      \mathbf{P}_{pp} & \mathbf{P}_{pd} \\
      \mathbf{P}_{dp} & \mathbf{P}_{dd}
    \end{pmatrix} \ .
  \end{equation}
$\mathbf{x}^{a}_{p}$ and $\mathbf{P}^{a}_{pp}$ are the updates of mean
vector and covariance matrix of the parameters at pilot point
locations, $\mathbf{x}^{a}_{d}$ and $\mathbf{P}^{a}_{dd}$ are the
corresponding quantities for the dynamic variables. %
$\mathbf{P}^{a}_{dp}$ and $\mathbf{P}^{a}_{pd}$ contain updated
covariances between pilot-point parameters and dynamic variables. %

Now we discuss the second step, i.e., the transfer of the update to
non-pilot point parameters by kriging interpolation. %
Through the kriging interpolation, the updates from Equations
(\ref{eq:ppkf-update-state}) and (\ref{eq:ppkf-update-cov}) are used
to compute updates for means and covariances of parameters at
non-pilot points: %
  \begin{equation}
    \label{eq:ppkf-interpolate-state}
    \mathbf{x}^{a}-\mathbf{x}^{f} \
    = P_{p}\left[
      \begin{pmatrix}
        \mathbf{x}^{a}_{p} \\
        \mathbf{x}^{a}_{d}
      \end{pmatrix}
      -
      \begin{pmatrix}
        \mathbf{x}^{f}_{p} \\
        \mathbf{x}^{f}_{d}
      \end{pmatrix}
    \right]
  \end{equation}
The interpolation operator $P_{p}$ is defined as
  \begin{equation}
    \label{eq:Pp-kf-1}
    P_{p}
    =
    \begin{pmatrix}
      1 & 0 \\ %
      \mathbf{P}_{rp}^{0}\mathbf{P}_{pp}^{-1} & 0 \\ %
      0 & 1 \\ %
    \end{pmatrix}
    \in R^{n_{s}\times (n_{p}+n_{d})} \ .
  \end{equation}
Here, $\mathbf{P}_{rp}^{0} \in R^{n_{r}\times n_{p}}$ is a fixed
covariance matrix between pilot point parameters and non-pilot point
parameters. %
The matrix $\mathbf{P}_{rp}^{0}$ is specified in advance. %
A suitable choice for $\mathbf{P}_{rp}^{0}$ are prior covariances
equal to the covariances used in the generation of the prior parameter
fields as these covariances reflect the prior knowledge. %
The interpolation equation for the covariance matrix is %
  \begin{equation}
    \label{eq:ppkf-interpolate-covariance}
    \mathbf{P}^{a}-\mathbf{P}
    = P_{p}\left[
      \begin{pmatrix}
        \mathbf{P}^{a}_{pp} & \mathbf{P}^{a}_{pd} \\
        \mathbf{P}^{a}_{dp} & \mathbf{P}^{a}_{dd}
      \end{pmatrix}
      -
      \begin{pmatrix}
        \mathbf{P}_{pp} & \mathbf{P}_{pd} \\
        \mathbf{P}_{dp} & \mathbf{P}_{dd}
      \end{pmatrix}
    \right]P_{p}^{T} \ .
  \end{equation}
A closer discussion of the interpolation operator $P_{p}$ can be found
in Section \ref{sec:special-covariance}. %

The equations for the PP-EnKF are very similar to Equations
(\ref{eq:ppkf-split}), (\ref{eq:ppkf-update-state}), and
(\ref{eq:ppkf-interpolate-state}) for the mean update of the PP-KF. %
The state vector is split for every realization in the ensemble: %
  \begin{equation}
    \label{eq:ppenkf-split}
    \mathbf{x}^{f}_{i} =
    \begin{pmatrix}
      \mathbf{x}^{f}_{p, i} \\ %
      \mathbf{x}^{f}_{r, i} \\ %
      \mathbf{x}^{f}_{d, i} %
    \end{pmatrix}
    \in R^{n_{s}} \ , \qquad i \in
    \left\{
      1, \cdots, n_{e}
    \right\} \ .
  \end{equation}
Then, the update of the state vector is calculated for each
realization of pilot point parameters and dynamic variables: %
  \begin{equation}
    \label{eq:ppenkf-update-state}
    \begin{pmatrix}
      \mathbf{x}^{a}_{p, i} \\
      \mathbf{x}^{a}_{d, i}
    \end{pmatrix}
    -
    \begin{pmatrix}
      \mathbf{x}^{f}_{p, i} \\
      \mathbf{x}^{f}_{d, i}
    \end{pmatrix}
    =
    \begin{pmatrix}
      \mathbf{P}_{pp, e} & \mathbf{P}_{pd, e} \\
      \mathbf{P}_{dp, e} & \mathbf{P}_{dd, e}
    \end{pmatrix}
    \mathbf{H}^{T} \left( \mathbf{P}_{yppy, e}
      + \mathbf{R} \right)^{-1}
    \left(
      \mathbf{d}_{i}-\mathbf{H}
      \begin{pmatrix}
        \mathbf{x}^{f}_{p, i} \\
        \mathbf{x}^{f}_{d , i}
      \end{pmatrix}
    \right)
    , \qquad i \in
    \left\{
      1, \cdots, n_{e}
    \right\}.
  \end{equation}
Finally, the interpolation is applied to each realization: %
  \begin{equation}
    \label{eq:ppenkf-interpolate-state}
    \mathbf{x}^{a}_{i}-\mathbf{x}^{f}_{i} \
    = P_{p}\left(
      \begin{pmatrix}
        \mathbf{x}^{a}_{p, i} \\
        \mathbf{x}^{a}_{d, i}
      \end{pmatrix}
      -
      \begin{pmatrix}
        \mathbf{x}^{f}_{p, i} \\
        \mathbf{x}^{f}_{d, i}
      \end{pmatrix}
    \right) \
    , \qquad i \in
    \left\{
      1, \cdots, n_{e}
    \right\}.
  \end{equation}
There are two main differences between the updates of the PP-EnKF and
the PP-KF. %
First, the update of the PP-EnKF consists of $n_{e}$ equations, one
for each realization of the ensemble. %
Second, all covariance matrices except $\mathbf{P}_{rp}^{0}$ are
calculated from the ensemble of state vector realizations. %
When moving from the PP-KF to the PP-EnKF, the initialization of the
parameter field is randomized. %
Opposed to this randomization, the variability outside the pilot
points, represented by the fixed matrix $\mathbf{P}_{rp}^{0}$, is
chosen once and then fixed throughout the computation. %
In the next section, important matrices of the PP-KF and the PP-EnKF
algorithms are explained in more detail. %

\subsection{Covariance matrix of the simulated observations and
  interpolation operator}
\label{sec:special-covariance}

The matrix $\mathbf{P}_{yppy}$ used in the update equations
(\ref{eq:ppkf-update-state}) and (\ref{eq:ppkf-update-cov}) of the
PP-KF is an approximation of the covariance matrix of simulated
measurement variables $\mathbf{P}_{yy}$ appearing in the Kalman update
(Equation (\ref{eq:kalman-update})): %
  \begin{equation}
    \label{eq:Pyy}
    \mathbf{P}_{yy}
    =
    \mathbf{H}
    \mathbf{P}
    \mathbf{H}^{T}
  \end{equation}
$\mathbf{P}_{yppy}$ should approximate $\mathbf{P}_{yy}$ by taking
into account only the dynamic variables and the pilot point
parameters. %
The straightforward approximation with this property is obtained by
removing the covariances that contain information from the non-pilot
point locations, i.e. $\mathbf{P}_{rp}$, $\mathbf{P}_{pr}$,
$\mathbf{P}_{rr}$, $\mathbf{P}_{dr}$, and $\mathbf{P}_{rd}$, from
$\mathbf{P}$ before applying $\mathbf{H}$. %
  \begin{equation}
    \label{eq:Pyppy}
    \mathbf{P}_{yppy}
    =
    \mathbf{H}
    \begin{pmatrix}
      \mathbf{P}_{pp} & 0 & \mathbf{P}_{pd} \\ %
      0 & 0 & 0\\ %
      \mathbf{P}_{dp} & 0 & \mathbf{P}_{dd}\\ %
    \end{pmatrix}
    \mathbf{H}^{T}
  \end{equation}
Note that $\mathbf{P}_{yppy}=\mathbf{P}_{yy}$ if $\mathbf{H}$ is a
linear operator depending only on pilot point parameters and dynamic
variables. %
A simple example of such a measurement operator $\mathbf{H}$ arises
when all measurement variables are either dynamic variables or
parameters at pilot points. %
For the PP-EnKF, the above remains true, with the amendment that the
covariance matrices
$\mathbf{P}_{yppy, e}, \mathbf{P}_{yy, e}, \mathbf{P}_{pp, e},
\mathbf{P}_{pd, e}, \mathbf{P}_{dp, e}$, and $ \mathbf{P}_{dd, e}$ are
calculated from an ensemble of realizations. %

Now the interpolation operator is discussed. %
The operator $P_{p}$ defines the kriging interpolation, the second
part of the update of both the PP-KF and the PP-EnKF. %
It determines the updates of non-pilot point parameters based on the
updates of pilot point parameters. %
The definition of $P_{p}$ was given in Equation \eqref{eq:Pp-kf-1},
and it makes use of two covariance matrices, $\mathbf{P}_{pp}$ and
$\mathbf{P}_{rp}^{0}$. %
The covariance matrix of the pilot point parameters $\mathbf{P}_{pp}$
is obtained from the forward simulation. %
Thus, the knowledge of $\mathbf{P}_{pp}$ originates from a mixture of
the initially drawn prior covariance matrix and of the filter
equations that were applied before the current update. %
In contrast, the covariance between the non-pilot point parameters and
the pilot point parameters $\mathbf{P}_{rp}^{0}$ is fixed throughout
the computation. %
The superscript zero emphasizes two characteristic features of
$\mathbf{P}_{rp}^{0}$. First, $\mathbf{P}_{rp}^{0}$ is fixed and,
second, its elements may be prior covariances. %
In our synthetic experiments, we construct the prior covariances
$\mathbf{P}_{rp}^{0}$ from the initial permeability field ensemble. %
The initial permeability field ensemble itself is obtained by
Sequential Gaussian Simulation using prior information in the form of
a semivariogram (SGSIM, \citet{Deutsch1992}). %
Fixing $\mathbf{P}_{rp}^{0}$ is an integral part of the PP-EnKF,
because this fixed covariance is one of the two reasons why we expect
the PP-EnKF to suppress spurious correlations between pilot point
parameters and non-pilot point parameters, the other reason being the
reduction of the number of parameters in the Kalman update. %

\subsection{Comparison to the classical EnKF and other EnKF methods}
\label{sec:comparison-to-enkf}

We compare the update equations (\ref{eq:ppenkf-update-state}) and
(\ref{eq:ppenkf-interpolate-state}) of the PP-EnKF to the update
equation (\ref{eq:enkf-update}) of the classical EnKF. %
If the measurement operator $\mathbf{H}$ operates on parameters at
non-pilot points, this particular measurement information will be lost
through the approximation in the PP-EnKF update, as shown in section
\ref{sec:special-covariance}. %
This is a hypothetical case, because one can always locate pilot
points at parameter measurement locations. %
If the measurement operator $\mathbf{H}$ operates only on parameters
at pilot points and dynamic variables, the main difference between the
update equations is the kriging interpolation of the PP-EnKF. %
$\mathbf{P}_{e}$ in the update equation (\ref{eq:enkf-update}) of the
EnKF can be partitioned the same way as in Equation
(\ref{eq:ppkf-split}): %
  \begin{equation}
    \label{eq:pe-split}
    \mathbf{P}_{e} =
    \begin{pmatrix}
      \mathbf{P}_{pp ,e} & \mathbf{P}_{pr ,e} & \mathbf{P}_{pd ,e} \\ %
      \mathbf{P}_{rp ,e} & \mathbf{P}_{rr ,e} & \mathbf{P}_{rd ,e}\\ %
      \mathbf{P}_{dp ,e} & \mathbf{P}_{dr ,e} & \mathbf{P}_{dd ,e}\\ %
    \end{pmatrix}
    \in R^{n_{s}\times n_{s}}.
  \end{equation}
The middle column of this matrix does not influence the update
equation, if the aforementioned restrictions on $\mathbf{H}$ hold. %
In the PP-EnKF, $\mathbf{P}_{e}$ is approximated by %
  \begin{equation}
    \label{eq:Pe-PP-enkf}
    P_{p}
    \begin{pmatrix}
      \mathbf{P}_{pp, e} & \mathbf{P}_{pd, e} \\
      \mathbf{P}_{dp, e} & \mathbf{P}_{dd, e}
    \end{pmatrix}
    = %
    \begin{pmatrix}
      1 & 0 \\ %
      \mathbf{P}_{rp}^{0}\mathbf{P}_{pp, e}^{-1} & 0 \\ %
      0 & 1 \\ %
    \end{pmatrix}
    \begin{pmatrix}
      \mathbf{P}_{pp, e} & \mathbf{P}_{pd, e} \\
      \mathbf{P}_{dp, e} & \mathbf{P}_{dd, e}
    \end{pmatrix}
    = %
    \begin{pmatrix}
      \mathbf{P}_{pp, e} & \mathbf{P}_{pd, e} \\ %
      \color{red}{\mathbf{P}_{rp}^{0}} &
      \color{red}{\mathbf{P}_{rp}^{0}\mathbf{P}_{pp,
          e}^{-1}\mathbf{P}_{pd, e}} \\ %
      \mathbf{P}_{dp, e} & \mathbf{P}_{dd, e} \\ %
    \end{pmatrix} \, .
  \end{equation}
It is of interest to compare the two columns of matrix
(\ref{eq:Pe-PP-enkf}) to the first and last column of $\mathbf{P}_{e}$
in Equation (\ref{eq:pe-split}). %
Differences between the two matrices appear in the second row. %
The left columns of these matrices determine the update coming from
measurements of parameters at pilot points. %
The updates of the EnKF and the PP-EnKF would be equal if the fixed
covariance matrix $\mathbf{P}_{rp}^{0}$ were equal to the ensemble
covariance $\mathbf{P}_{rp, e}$ from the EnKF. %
Of course, the matrices will never be exactly equal due to sampling
fluctuations. %
One of the positive effects of $\mathbf{P}_{rp}^{0}$ is that it does
not suffer from spurious correlations. %

The right column of matrix (\ref{eq:Pe-PP-enkf}) determines the update
coming from measurements of dynamic variables. %
Values at non-pilot points are updated according to
$\mathbf{P}_{rp}^{0}\mathbf{P}_{pp, e}^{-1}\mathbf{P}_{pd, e}$ instead
of $\mathbf{P}_{rd, e}$. %
There is no direct covariance matrix between non-pilot points and
dynamic variables in the PP-EnKF. %
Instead, the update is correlated with the pilot points and then
interpolated according to $\mathbf{P}_{rp}^{0}$. %
Again, the main enhancement is the reduction of spurious
correlations. %

We now compare the update equations of the PP-EnKF to the update
equations of two popular EnKF methods, the local EnKF and the hybrid
EnKF. %
Similarly to the local EnKF and the hybrid EnKF, the PP-EnKF uses a
partially fixed covariance structure in order to suppress spurious
correlations. %
In the local EnKF, updates are calculated directly for all parameters,
but correlations between parameters (or correlations between dynamic
variables and parameters) are set to zero if their distance to
measurement locations exceeds a certain threshold. %
Like the local EnKF, the PP-EnKF aims to suppress spurious
correlations. %
To achieve this, only the correlations between measurement variables
and pilot points are calculated from the ensemble. %
We argue that the PP-EnKF could be preferred to the local EnKF in
cases, when distant locations are significantly correlated. %
Correlations between these locations would be suppressed by the local
EnKF, while the PP-EnKF includes them through correlations between
distant pilot point locations. %

In the hybrid EnKF, a mixture of covariance matrices is used in the
update step, partially fixed and partially calculated from the
ensemble. %
While this can lead to good results, we argue that the PP-EnKF
delivers an appealing alternative, because it interferes less with the
statistics of the update step. %
The first step of the update of the PP-EnKF is restricted to pilot
point parameters and dynamic variables. %
As a benefit, the update for these two sets of state vector variables
is calculated exclusively from the ensemble-based correlations that
are derived from the ensemble of forward model runs. %
Only the rest of the parameters is then updated by kriging the updates
at the pilot points, using the pre-defined fixed covariance matrix. %
For the hybrid EnKF, all updates are, at least partially, subject to
the pre-defined, fixed covariance matrix. %

An additional benefit of the PP-EnKF is that its implementation is
relatively straightforward. %
In contrast to the hybrid or the local EnKF, there is no need to
change the covariance matrix of the classical EnKF update equation. %
All changes of the PP-EnKF (compared to the classical EnKF) can be
implemented by pre-processing (restricting the state vector to dynamic
variables and parameters at pilot points) and post-processing
(calculating the interpolated parameter values at non-pilot points). %
Thus, a modular implementation around an existing implementation of
the EnKF update is possible. %

\subsection{Optimality considerations for the PP-EnKF}
\label{sec:optimality-considerations}

Finally, in this section we discuss the optimality of the PP-EnKF. %
To this end, we consider the EnKF as the benchmark optimal method,
even though it has its own limitations \citep{Evensen2003}. %
All filters in the comparisons described in this paragraph are
considered to be applied in combination with an infinite ensemble,
thus without any spurious correlations. %
For a given use case, two comparisons will be discussed between the
PP-EnKF and two EnKF variants that will serve as optimality bounds
(see also Table \ref{tab:optimality-bounds}). %
The given use case can be thought of as the model setups described in
this work, but the reasoning in this paragraph holds for any model
setup where pilot points can be applied. %

The first comparison is between the PP-EnKF and the EnKF, both applied
in a synthetic data assimilation experiment for the given use case. %
By definition of its update, the PP-EnKF is sub-optimal to the EnKF
for a given use case under perfect conditions, i.e., at the limit of
infinite ensemble size. %
This is because the update of the PP-EnKF is interpolated while the
update of the EnKF retains the full variability of the parameters in
the state vector. %

We now turn to the second comparison between the PP-EnKF and an
altered EnKF, from now on called Interpolated EnKF. %
Again, both filters are applied in two synthetic data assimilation
experiments for the same use case. %
The Interpolated EnKF uses a model setup that only consists of pilot
point parameters and dynamic states. %
In the Interpolated EnKF, the full set of parameters is interpolated
from the pilot points for both the updates and the forward
computation. %
In our PP-EnKF, in contrast, the full set of parameters is used for
the forward computation, while the updates are interpolated from the
pilot points as in the Interpolated EnKF. %
It follows that the Interpolated EnKF is sub-optimal to the PP-EnKF,
since it uses interpolated values in the forward computation, where
the PP-EnKF retains the full variability of the (infinite and
correct) initial parameter ensemble. %
On the other hand, the Interpolated EnKF is still optimal on its own
right when defining the pilot points to be the only parametric
degrees of freedom of the same use case. %

In summary, the PP-EnKF is theoretically sub-optimal to the EnKF,
while the Interpolated EnKF is sub-optimal to the PP-EnKF, although
optimal under purely pilot point based parametrization. %
This way, we use the EnKF and the Interpolated EnKF as optimality
bounds for the PP-EnKF. %
As a final remark, when realistic models and limited ensemble sizes
are considered, the PP-EnKF regains its advantage over the EnKF
regarding the suppression of spurious correlations that we
illustrate in this study. %

\begin{table}[h]
  \centering
  \caption{Optimality bounds for the PP-EnKF. %
    The PP-EnKF is compared to the EnKF and to the Interpolated EnKF
    that is defined in the text. \\ %
  } %
  \label{tab:optimality-bounds}
  \begin{tabular}{l | c c c}
    full variability in... 
    &
      EnKF
    &
      PP-EnKF
    &
      Interpolated EnKF
    \\
    \hline
    forward computation
    &
      \checkmark
    &
      \checkmark
    &
      x
    \\
    update equation
    &
      \checkmark
    &
      x
    &
      x
    \\
    \hline
  \end{tabular}
\end{table} %

\section{Design of the synthetic experiments}
\label{sec:design-syn-exp}

The performance of the PP-EnKF is evaluated by comparing it to seven
other EnKF methods (damped, iterative, local, hybrid, dual
\citep{Moradkhani2005}, normal score \citep[NS-EnKF][]{Zhou2011_3,
  Schoniger2012, Li2012, Goovaerts1997, Journel1978} and classical). %
The comparison procedure is adopted from \citet{Keller2018}. %
As in \citep{Keller2018}, the performance evaluations are carried out
for two parameter estimation setups: a 2D tracer transport problem and
a 2D flow problem with one injection well. %
The estimated parameter field in these setups is the permeability
field that largely determines the hydraulic conductivity field in the
subsurface. %
In each setup, $1,000$ synthetic experiments are computed for the
PP-EnKF and for the seven other EnKF methods, and each one for
ensemble sizes of 50, 70, 100, and 250. %
In each experiment, we assess both the updated parameter fields and
their ensemble variance. %
The ensemble variance is compared to a reference EnKF estimation with
a very large ensemble size of 10,000. %
Next to the RMSE comparison, there is a comparison of the correlation
fields driving the Kalman update using 10 synthetic experiments for
ensemble size 50. %
The correlation output of these synthetic experiments is compared to
the correlation field of the reference EnKF with ensemble size 10,000
by computing the RMSE difference of the correlation fields. %
Additionally, the impact of the correlation length of the prior
permeability fields is investigated. %
The correlation length is important in the PP-EnKF, because it not
only affects the prior permeability field but also the kriging
interpolation. %
Finally, to check the influence of the pilot point grid on the
performance of the PP-EnKF, different grid configurations are
tested. %

\subsection{Subsurface models}
\label{sec:subsurface-models}

\begin{figure}
  \centering
  \includegraphics[trim = 50 100 100 80, clip,
    width=\columnwidth]{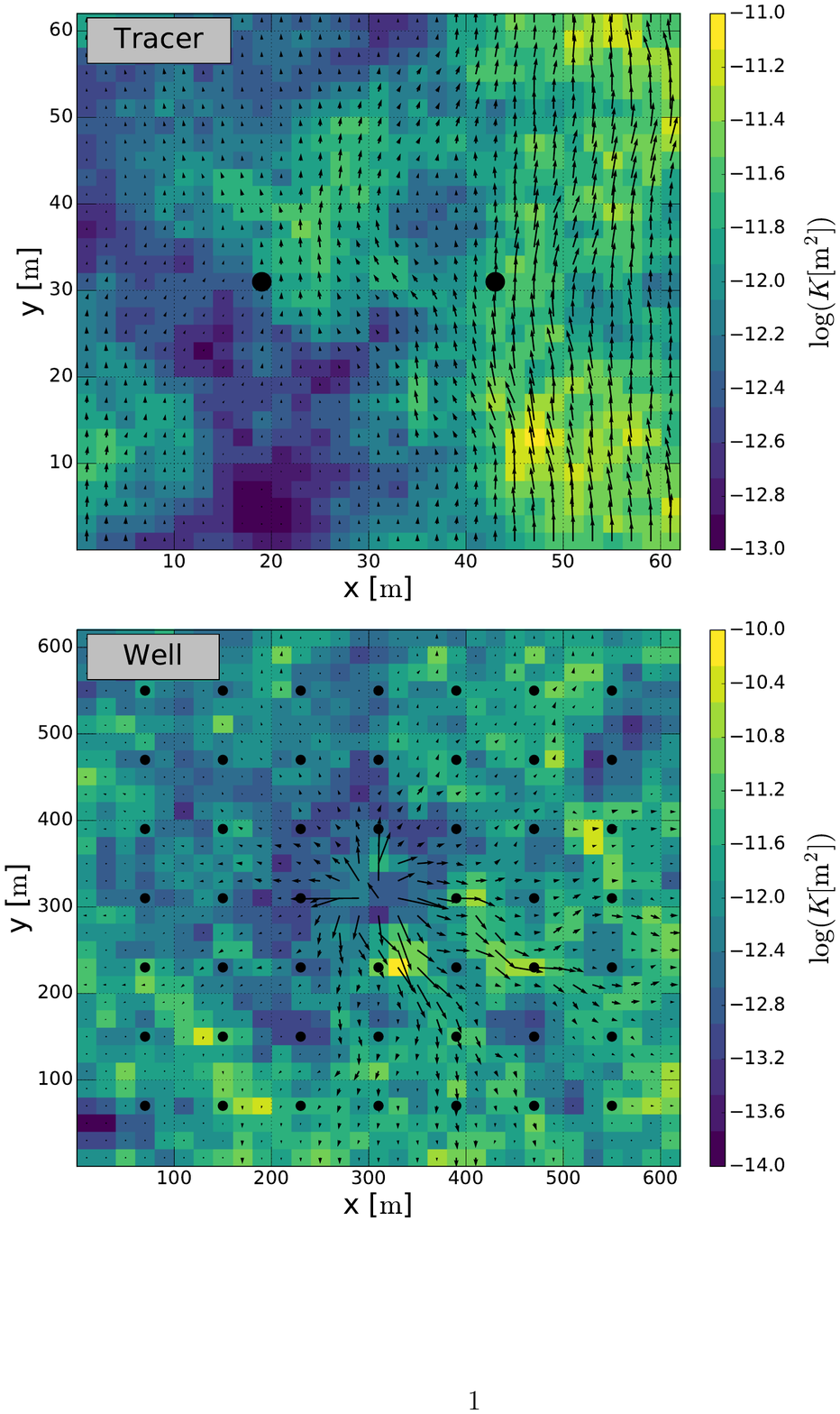}
    \caption{The logarithmic reference permeability fields for the
      tracer model and the well model with groundwater flow vectors.
      Measurement locations are depicted as black circles. Figure
      taken from \citet{Keller2018}.}
  \label{fig:true-perm}
\end{figure}%

The performance comparison of the PP-EnKF with other EnKF-variants in
this study is carried out for two transient subsurface model setups,
first, a 2D solute transport model (tracer model), and, second, a 2D
groundwater flow problem with an injection well and four pumping wells
(well model). %
These model setups have already been used in a previous publication
that compared existing EnKF methods without the PP-EnKF and are used
as benchmark for the PP-EnKF here \citep{Keller2018}. %
The grids of both models consist of $31\times31 = 961$ identical
squared cells. %
The size of the model domain for the tracer model is
$62\, \mathrm{m} \times 62\, \mathrm{m}$; the size of the model domain
for the well model is $620\, \mathrm{m} \times 620\, \mathrm{m}$. %
The simulation period is $1,200$ days for the tracer model and $18$
days for the well model. %
For both models, the simulation period is divided into $1,200$ time
steps. %
All forward simulations are computed with the numerical software
SHEMAT-Suite \citep{Rath2006,Clauser2012,Keller2020}. %

In the tracer model, a constant head difference of
$11\, \mathrm{m}- 10\, \mathrm{m} = 1\, \mathrm{m}$ is implemented
between the southern boundary and the northern boundary, accompanied
by a constant concentration difference of
$80\times10^{-3}\, \mathrm{mol}/\mathrm{L}-60\times10^{-3}\,
\mathrm{mol}/\mathrm{L} =20\times10^{-3}\, \mathrm{mol}/\mathrm{L}$. %
The remaining two boundaries are impermeable. %
The initial conditions are a head of $10\, \mathrm{m}$ and a
concentration of $60\times10^{-3}\, \mathrm{mol}/\mathrm{L}$
throughout the model domain for the tracer model. %
In the well model, a head difference of $1\, \mathrm{m}$ is
implemented between a central injection well location at coordinates
($310\, \mathrm{m}$, $310\, \mathrm{m}$) and the four boundaries. %
Initial head is chosen as $10\, \mathrm{m}$ throughout the model
domain for the well model. %
For both setups, standard properties of water are used and the
porosity of the background matrix is $10\%$. %
The tracer is subjected to advective transport only. %

Measurement locations vary between the two setups. %
The tracer model includes two measurement locations at coordinates
$\left(19\, \mathrm{m},31\, \mathrm{m}\right)$ and
$\left(43\, \mathrm{m},31\, \mathrm{m}\right)$. %
At these locations, tracer concentration and hydraulic head
measurements are available at $100$ evenly distributed times
throughout the simulation period. %
The well model includes 49 measurement locations on a $7\times 7$ grid
throughout the model domain. %
At these locations, hydraulic head measurements are available at $60$
evenly distributed times throughout the simulation period. %

The synthetic reference permeability distributions for the two
subsurface model setups are displayed in Figure \ref{fig:true-perm}. %
These permeability fields are generated by sequential multi-Gaussian
simulation (SGSIM, \citet{Deutsch1992}). %
The same holds for the ensemble of prior permeability field
realizations that is used to initialize the EnKF. %
A permeability mean of $-12.0\, \log_{10}(K[\mathrm{m}^{2}])$ is used
for the synthetic reference and a permeability mean of
$-12.5\, \log_{10}(K[\mathrm{m}^{2}])$ is used for the prior
permeability distributions. %
In both cases, the standard deviation is
$0.5\, \log_{10}(K[\mathrm{m}^{2}])$. %
The isotropic correlation length of the permeability fields is
$50\, \mathrm{m}$ for the tracer model and $60\, \mathrm{m}$ for the
well model. %
No nugget effect is used. %
More details on the spherical correlation function used can be found
in \citet{Keller2018}. %
Additionally, the input files for the sequential Gaussian simulation
can be found in the data repository of this publication. %

\subsection{Parameter settings for the PP-EnKF and other
  EnKF variants}
\label{sec:design-pp-enkf}

Measurement noises are equal for all EnKF methods including the
PP-EnKF. %
Measurement noises for the concentration measurements of the tracer
model are set to
$\sigma_{\mathtt{c}}= 7.1\times 10^{-3}\mathrm{mol}/\mathrm{L}$ and
measurement noises for the hydraulic head measurements of both setups
are set to $\sigma_{\mathtt{h}}= 5\times 10^{-2}~\mathrm{m}$. %
EnKF methods which require parameter settings include the damped EnKF,
where the damping constant is set to $0.1$ \citep{Hendricks2008}. %
In the local EnKF, the length scale (which is half the cutoff radius)
is set to $150\, \mathrm{m}$, which is larger than the correlation
lengths. %
In the hybrid EnKF, the mixing constant is set to $0.5$ and a diagonal
background covariance matrix is specified. %
The damping factor of the damped EnKF and the parameter choices of the
local EnKF and the hybrid EnKF were shown to yield the smallest root
mean square errors in most synthetic experiments of the two setups of
a previous performance comparison \citep{Keller2018}. %

The PP-EnKF needs as inputs the locations for the pilot points, and a
covariance matrix for the kriging interpolation. %
We use 51 pilot points including all cell indices of the measurement
locations of both setups (tracer model and well model). %
Thus, the pilot points lie on a $7\times 7$ grid for both model setups
with two additional pilot points in the center of the model
corresponding to the measurement locations of the tracer model. %
A regular grid of pilot points is documented to be beneficial
\citep{Capilla1997}. %
The interpolation covariance $\mathbf{P}^{0}_{rp}$ is chosen identical
to the prior covariance. %
This is implemented by computing the covariances in
$\mathbf{P}^{0}_{rp}$ from 10,000 permeability fields, which are
generated by SGSim with the same correlation length, mean and standard
deviation as the prior permeability fields. %

\subsection{Performance Comparison Setup}
\label{sec:perf-comp-setup}

In \citet{Keller2018}, seven EnKF variants have been compared (damped,
iterative, local, hybrid, dual, normal score and classical). %
Now, we compare the PP-EnKF to these same seven methods by computing
synthetic experiments for the two physical model setups introduced in
the previous sections (compare Section
\ref{sec:comparison-subsection}). %
Additionally, correlation lengths of the initial permeability fields
of both setups are varied to half and twice the correlation length of
the synthetic truth (Section
\ref{sec:comparison-correlation-lengths}). %
This is done, because the correlation length plays a very prominent
role in the PP-EnKF, more prominent than in the other EnKF methods,
since it enters in the PP-EnKF not only as correlation length of the
prior permeability field, but also as correlation length of the
kriging interpolation. %
Thus, it is especially important to see how the PP-EnKF performs
compared to other models, when it is subject to mis-specified
correlation lengths. %
One important degree of freedom in the PP-EnKF is the choice of the
grid of pilot points. %
Thus, we compare the results of different sensible grid choices for
the two model setups (Section \ref{sec:var-pp-grids}). %
Next to the performance of the mean update, we are interested in the
spatial variability of the permeability ensemble update by the
PP-EnKF. %
This will help assess the ability of the PP-EnKF to suppress spurious
correlations. %
To evaluate the reproduction of the uncertainty of the estimates, we
compare the overall standard deviation of the ensembles generated by
the updates of the various EnKF methods. %
For each EnKF method, 1,000 synthetic experiments are used. %
The synthetic experiments differ solely in their random seed for
geostatistical ensemble initialization and measurement perturbation. %
Additionally, the full correlation fields of the EnKF and the PP-EnKF
from 10 synthetic experiments are compared to a reference correlation
field. %

\subsection{Performance Comparison Measures}
\label{sec:comparison-method}

Multiple synthetic experiments are needed to compare EnKF methods
\citep{Keller2018}. %
Here we use root-mean-square errors (RMSEs) from 1,000 synthetic
experiments to compare the PP-EnKF to the other seven EnKF methods. %
For each synthetic experiment $j$, the RMSE is computed as follows
  \begin{equation}
    \mathrm{RMSE}_{j}  = \sqrt{
      \frac{1}{n_{g}}\sum_{i = 1}^{n_{g}}
      \left(
        \bar{Y}_{i, j}-Y^{t}_{i, j}
      \right)^{2}
    }, \qquad j \in
    \left\{
      1, \cdots, 1000
    \right\}.
    \label{eq:rmse}
  \end{equation}
Such a single RMSE is computed from the squared differences of
estimated mean logarithmic permeabilities
$\bar{\mathbf{Y}}_{j} \in R^{n_{g}}$ and the synthetic reference
$\mathbf{Y}^{t}_{j} \in R^{n_{g}}$ across the $n_{g}$ grid cells. %

The RMSE measures the distance between the average over the estimated
permeability field realizations and the synthetic true permeability
field. %
Additionally, the overall standard deviation among realizations is
introduced as a measure for the uncertainty of the estimated
permeability field realizations. %
It is called STD in the remainder of this text. %
This overall standard deviation for a single synthetic experiment $j$
is calculated as the square root of the mean over the domain of the
pixel-wise ensemble variances, as follows
  \begin{equation}
    \mathrm{STD}_{j}  = \sqrt{
      \frac{1}{n_{g}}\sum_{i = 1}^{n_{g}}
      \sigma_{i,j}^{2}
    }, \qquad j \in
    \left\{
      1, \cdots, 1000
    \right\}.
    \label{eq:std}
  \end{equation}
where $n_{g}$ is the number of grid cells. %
The sample variances in Equation \eqref{eq:std} are calculated as
follows
  \begin{equation}
    \sigma_{i,j}^{2} =
    \frac{1}{n_{e}-1}\sum_{k=1}^{n_{e}}
    \left(
      Y_{k, i, j} - \bar{Y}_{i, j}
    \right)^{2}
    \label{eq:sample-var}
  \end{equation}
where $n_{e}$ is the number of realizations in the ensemble,
$\bar{Y}_{i,j}$ is the mean logarithmic permeability for synthetic
experiment $j$ and grid cell $i$, and $Y_{k,i,j}$ is the permeability
of realization $k$ calculated at grid cell $i$ in synthetic experiment
$j$. %

Besides the overall performance assessment combining results of the
RMSE and STD, a smaller test with 10 synthetic experiments is carried
out for checking the correlations between observed variables and the
full parameter field. %
This input/output-intensive check is executed only for the PP-EnKF and
the classical EnKF. %
A reference synthetic experiment is computed using the EnKF with an
ensemble size of 10,000. %
Subsequently, 10 synthetic experiments were computed for EnKF and
PP-EnKF using ensemble size 50. %
For the tracer model, the correlations are of the form
  \begin{equation}
    \rho(c_{obs},K) =
    \frac{\mathrm{Cov}(c_{obs},K)}{\sigma_{c}\cdot\sigma_{K}}
    \label{eq:corr-conc}
  \end{equation}
where the covariance $\mathrm{Cov}$ and the standard deviations
$\sigma$ are estimated from the ensemble, $c_{obs}$ denotes the
concentration at one of the two observation locations, and $K$ denotes
the logarithmic permeability at any given location in the field. %
For the well model, the correlations are of the form
  \begin{equation}
    \rho(h_{obs},K) =
    \frac{\mathrm{Cov}(h_{obs},K)}{\sigma_{h}\cdot\sigma_{K}}
    \label{eq:corr-head}
  \end{equation}
where the difference to the tracer case is that $h_{obs}$ denotes the
head observed at one of the 49 measurement locations. %
By varying $K$ across the model domain, we obtain the field of
correlations for each synthetic experiment. %
The RMSEs between these correlation fields and the reference
correlation field is used to judge the amount of spurious correlation
in the synthetic experiments introduced by the small ensemble size. %

\section{Results}
\label{sec:perf-comp}

\subsection{Comparison of the PP-EnKF to other EnKF methods}
\label{sec:comparison-subsection}

\begin{figure}
  \centering
  \includegraphics[trim = 50 200 50 150, clip,
    width=\columnwidth]{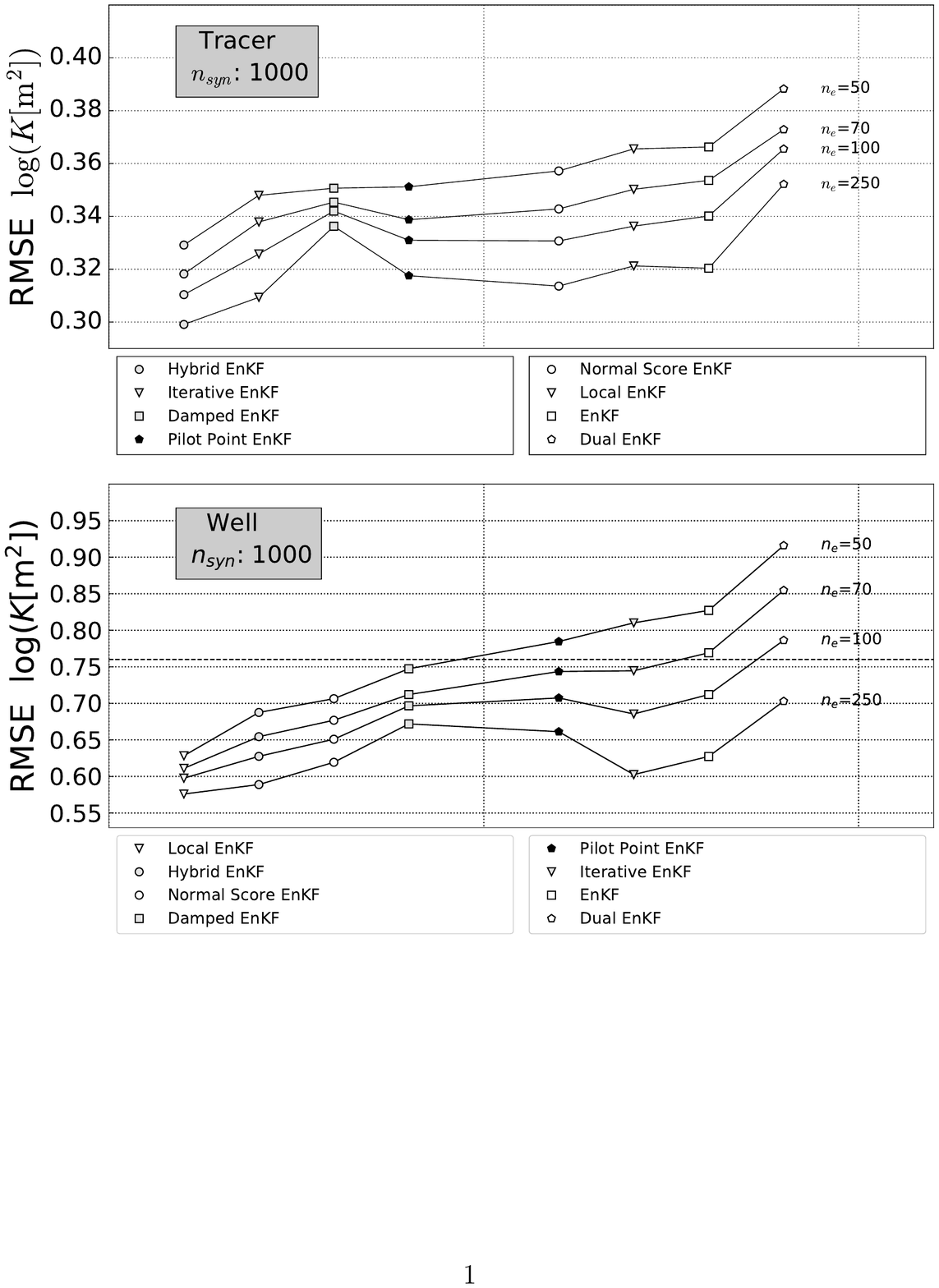}
    \caption{Comparison of the root mean square errors (RMSEs) of the
      pilot point EnKF and other EnKF methods in the tracer model
      (top) and well model (bottom). %
    }
    \label{fig:model-comp}
\end{figure}%

To assess its performance, we compare the PP-EnKF to seven other EnKF
methods. %
First, results for the tracer model setup are shown. %
In Figure \ref{fig:model-comp} (top), the mean RMSEs from 1,000
synthetic experiments are shown for the eight EnKF methods. %
Numerical results are given in the supporting information. %
The tracer model is known to yield a small range of RMSEs across the
tested EnKF methods \citep{Keller2018}. %
Among the eight tested EnKF methods, the PP-EnKF ranks among the four
methods with the smallest RMSEs for all tested ensemble sizes. %
For ensemble size 50 and 250, the PP-EnKF has the fourth smallest
RMSE. %
For ensemble sizes 70 and 100, it yields the third smallest RMSE. %

A direct comparison to the classical EnKF is interesting for two
reasons. %
First, the PP-EnKF is directly derived from the classical EnKF. %
Second, earlier publications include conflicting results regarding
this comparison. %
In the synthetic experiments performed with our tracer model, the
PP-EnKF performs significantly better than the classical EnKF for
ensemble sizes 50 70, and 100, and slightly better for ensemble size
250. %

Turning to results from the well model setup, Figure
\ref{fig:model-comp} (bottom) shows the RMSE means from the
corresponding 1,000 synthetic experiments. %
Numerical results are given in the supporting information. %
Compared to the tracer model, the well model yields larger RMSE
differences between the tested methods. %
The PP-EnKF yields slightly worse than average RMSEs for the well
model. %
For ensemble size 50 and 70, the PP-EnKF has the fifth smallest
RMSE. %
For ensemble sizes 100, and 250, it scores the sixth smallest RMSE. %
Compared to the classical EnKF, the PP-EnKF performs better for
ensemble sizes 50, 70, slightly better for ensemble size 100, and
worse for ensemble size 250. %

Summing up the results for the two physical model setups, the PP-EnKF
yields good RMSE results for the tracer model and medium RMSE results
for the well model. %
It should be noted that these setups were taken from our previous
study, and that a very robust testing was done using $1,000$ synthetic
studies. %
Results from both physical model setups suggest that, for small
ensemble sizes, the PP-EnKF is a clear improvement to the classical
EnKF as desired by design. %
This is in slight contradiction to results from \citet{Heidari2013},
where the PP-EnKF yields larger RMSEs than the classical EnKF for a
synthetic experiment with ensemble size 50. %
Our results for ensemble size 250, for which both methods yield
similar results, are in agreement with similar results from
\citet{Tavakoli2013}. %

Furthermore, the RMSEs suggest that the PP-EnKF provides a trade-off
between damping and the normal EnKF. %
For ensemble size 50, the PP-EnKF has similar or larger RMSE than the
damped EnKF, but smaller RMSE than the classical EnKF. %
Moving to larger ensemble sizes, the reduction of RMSE by the PP-EnKF
is much larger than the reduction of RMSE by the damped EnKF. %
A possible explanation for this effect is that the interpolated
updates of the PP-EnKF reduce spurious correlations and thereby reduce
the number of divergent synthetic experiments for small ensemble
sizes. %
For larger ensemble sizes, the relatively large reductions of the
RMSEs of the PP-EnKF suggest that the interpolated updates of the
PP-EnKF can incorporate more information than the damped updates of
the damped EnKF.

The full RMSE distributions give a more in-depth picture of the
performance of the EnKF methods. %
They are provided in the supporting information. %
The distributions illustrate how the RMSE means in the previous
section were obtained. %
For the well model, one can see that the PP-EnKF yields a somewhat
wider RMSE distribution than other methods. %

\subsection{Ensemble variance of conditional realizations}
\label{sec:ensemble variance}

Now, we test and compare the uncertainty characterization of
conditional realizations obtained from the PP-EnKF by looking at the
overall ensemble standard deviations. %
The equations for the overall standard deviations were introduced in
Section \ref{sec:comparison-method}. %
We compare the mean overall standard deviations calculated over 1,000
synthetic experiments for each EnKF method. %
Additionally, a benchmark STD is derived from 100 synthetic
experiments using the classical EnKF with ensemble size 10,000. %

\begin{figure}
  \centering
  \includegraphics[trim = 50 190 50 150, clip,
    width=\columnwidth]{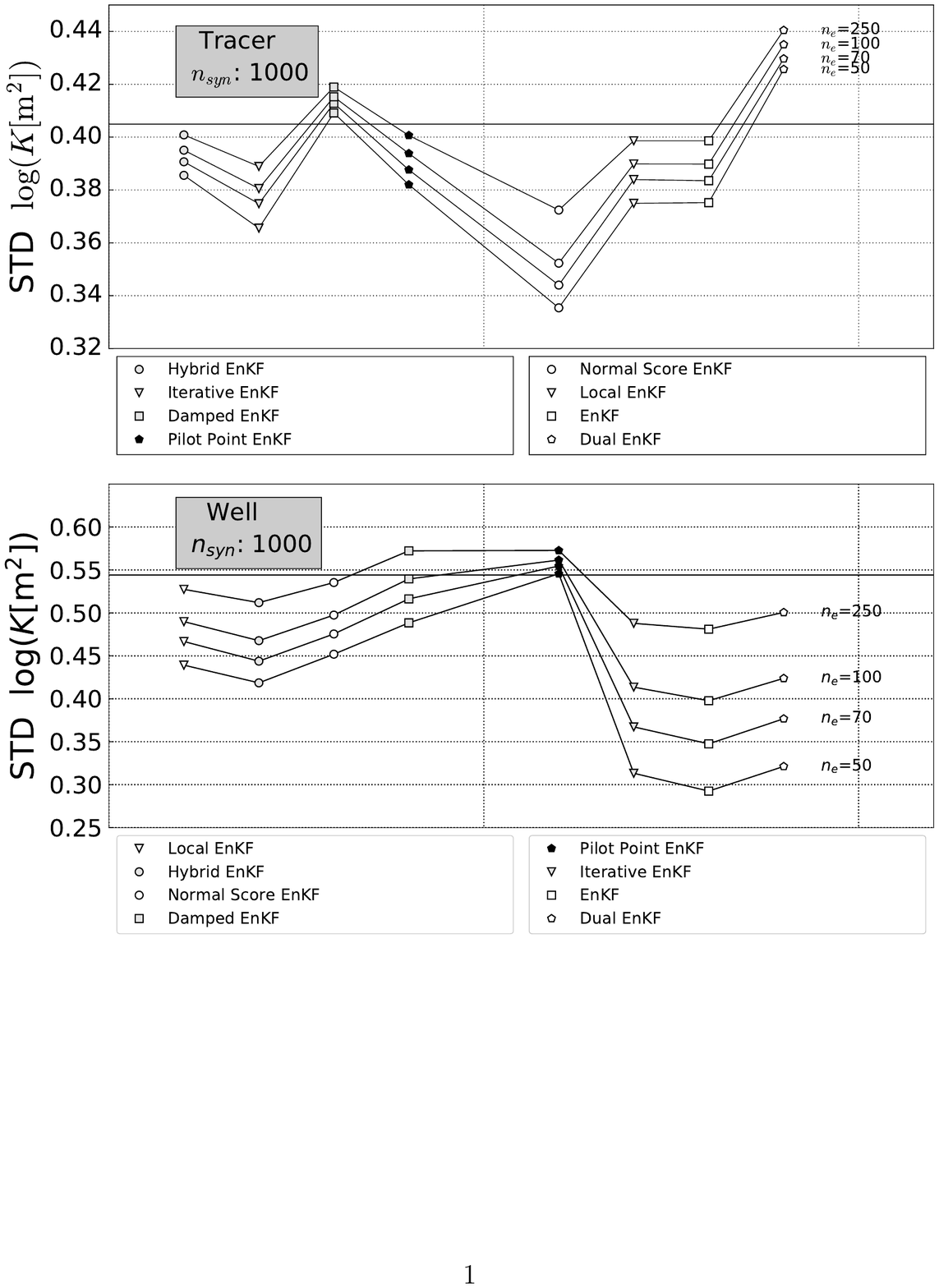}
    \caption{Comparison of the overall standard deviation (STD) of the
      pilot point EnKF and other EnKF variants for the tracer model
      (top) and well model (bottom). The horizontal black line depicts
      the STD obtained from 100 synthetic experiments with 10,000
      ensemble members. %
    }
  \label{fig:mean-std-comps}
\end{figure}%

The results for the tracer model are shown in the top half of Figure
\ref{fig:mean-std-comps}. %
All EnKF methods yield standard deviations that are reasonably close
to the standard deviation of the benchmark run. %
Still, most methods underestimate the ensemble spread; exceptions are
the damped EnKF and the dual EnKF. %
Comparing the differences to the 10,000 ensemble run, the pilot point
EnKF ranks fourth for ensemble size 50, third for ensemble sizes 70
and 100, and second for ensemble size 250. %
Thus, the pilot point EnKF is in the top half of the methods, not only
for RMSE comparison, but also concerning the standard deviation. %

For the well model, the mean overall standard deviations are shown in
the bottom half of Figure \ref{fig:mean-std-comps}. %
Compared to the tracer model, there is a larger tendency of most
methods to underestimate ensemble spread compared to the 10,000
ensemble run. %
Additionally, one can see a clear difference between the EnKF methods
that contain a form of damping and the ones that do not (including the
iterative, classical and dual EnKF). %
The pilot point EnKF yields a large ensemble spread, especially for
ensemble size 50, where it yields the best STD of all methods. %
For ensemble size 70, the pilot point EnKF still ranks first, for
ensemble size 100 it ranks second and for ensemble size 250 it ranks
third. %
Thus, while the pilot point EnKF only obtained medium results in RMSE
comparison compared to other EnKF methods, it always ranks in the top
three methods concerning uncertainty characterization, ranking first
for the important smallest ensemble sizes.

In conclusion, the uncertainty characterization of the pilot point
EnKF is generally good compared to other EnKF methods for the tracer
model and well model. %
This is one of the main features of the method, and this benefit again
comes by design. %
In the update step, the erroneous reduction of ensemble variance is
constrained by two effects. %
First, by reducing the number of parameters to the number of
parameters at pilot points, issues of rank deficiency and inbreeding
are reduced, and second, by interpolating the update, large parts of
the prior variability remain intact. %
Our results agree with results from \citet{Heidari2013}, where a large
spatial variability in PP-EnKF updates is diagnosed. %
While \citet{Tavakoli2013} also diagnose heterogeneities in the
PP-EnKF results, the small ensemble spread in their PP-EnKF results is
contradictory to our findings. %
This might be due to the specific subsurface model in
\citep{Tavakoli2013}, where only a small fraction of the model
parameters was updated. %

\subsection{Variation of prior correlation lengths}
\label{sec:comparison-correlation-lengths}

In this section, results for erroneous correlation lengths are
discussed: the prior correlation lengths are varied to half or twice
the correlation length of the synthetic truth. %
These results are especially important for judging the performance of
the PP-EnKF, since it is influenced by the correlation length through
both the prior realizations and the interpolation. %
The numerical RMSE results discussed in this section can be found in
the supporting information. %

\begin{figure}
  \centering
  \includegraphics[trim = 50 190 50 150, clip,
    width=\columnwidth]{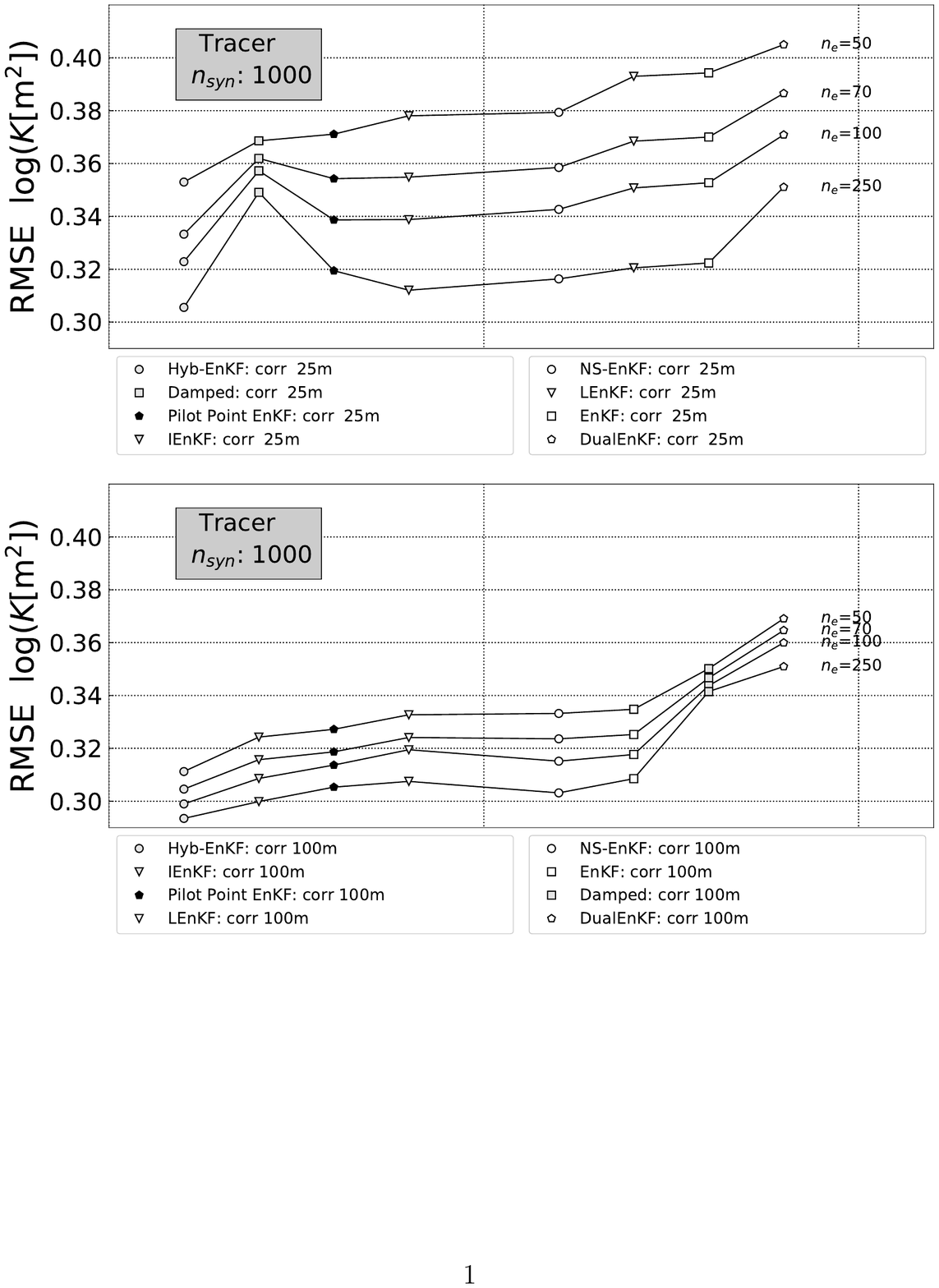}
    \caption{Comparison of the pilot point EnKF with other
      EnKF variants for the tracer model and correlation lengths
      $25\,\mathrm{m}$ (top) and $100\,\mathrm{m}$ (bottom). %
    }
  \label{fig:tracer-model-corr-comp}
\end{figure}%

First, we discuss the tracer model with correlation length
$25\,\mathrm{m}$. %
This correlation length is half the correlation length of the
synthetic truth. %
Figure \ref{fig:tracer-model-corr-comp} (top) shows the average RMSE
values comparing to the results in Section
\ref{sec:comparison-subsection}. %
For ensemble size 50, all EnKF methods yield larger RMSEs than for the
standard case. %
For ensemble size 250, all methods except the classical EnKF and the
dual EnKF yield larger RMSEs than for the standard case. %
A specific look at the PP-EnKF shows that, for ensemble size 50, it
ranks third among the EnKF methods. %
For ensemble sizes 70 and 100, it ranks second, and for ensemble size
250, third. %
Relatively to the other EnKF methods, these results are a slight
improvement for the PP-EnKF compared to the case of the correct
correlation length. %
This means that the PP-EnKF seems to be robust against a too small
specification of correlation length for this test case. %

Now we discuss the results for the tracer model and a too long
correlation length of $100\,\mathrm{m}$ as shown in Figure
\ref{fig:tracer-model-corr-comp} (bottom). %
For all ensemble sizes, all methods except the damped EnKF yield
smaller RMSEs than for the correct correlation length
($50\,\mathrm{m}$). %
The damped EnKF yields slightly larger RMSEs for ensemble sizes 100
and 250 than for the standard case. %
Regarding the PP-EnKF, it ranks third among all methods for ensemble
sizes 50, 70 and 100. %
For ensemble size 250, it ranks fourth among all methods. %
Thus, the ranking of the PP-EnKF among the different methods is
comparable to the corresponding results for the correct correlation
length. %

For a discussion of these results, recall how an erroneous correlation
length may affect the performance of EnKF methods. %
The correlation length primarily affects the prior permeability
fields. %
This should generally be a disadvantage for the update, thus leading
to larger RMSEs. %
An additional effect of the correlation length is its direct influence
on the EnKF update. %
The correlations that drive the update are either more restricted to
the vicinity of the measurement locations, for the case of a small
correlation length, or they are spread out more widely for the case of
the large correlation length. %
The RMSE result for the small correlation length suggest that the
restriction of the EnKF update is particularly obstructive for the
tracer model, possibly because there are only two measurement
locations in this setup. %
Thus, for a small number of measurement locations, an underestimation
of the correlation length may significantly inhibit EnKF updates. %
On the other hand, the smaller RMSEs for correlation length
$100\,\mathrm{m}$ are surprising. %
This result suggests that the effect of the larger update radius is
stronger than the effect of the wrong prior correlation length in the
tracer setup. %
Good results for too large correlation lengths have been documented in
the literature, for example by \citet{Chaudhuri2018}. %
\citet{Chaudhuri2018} attribute this effect to two possible factors. %
First, too small correlation lengths may not capture the full
information content of the observations, and second, the too long
correlations lengths may lead to smoothed fields that prevent the
appearance of locally strong deviations from the true reference
field. %
However, the beneficial effect of long correlations lengths may vanish
for more complicated synthetic reference fields, especially, when they
exhibit small-scale heterogeneities. %
For example, in \citet{Camporese2011, Camporese2015} it was found that
an overestimated prior correlation length can propagate spurious
correlations, leading to worse results than an underestimated prior
correlation length. %

\begin{figure}
  \centering
  \includegraphics[trim = 50 190 50 150, clip,
    width=\columnwidth]{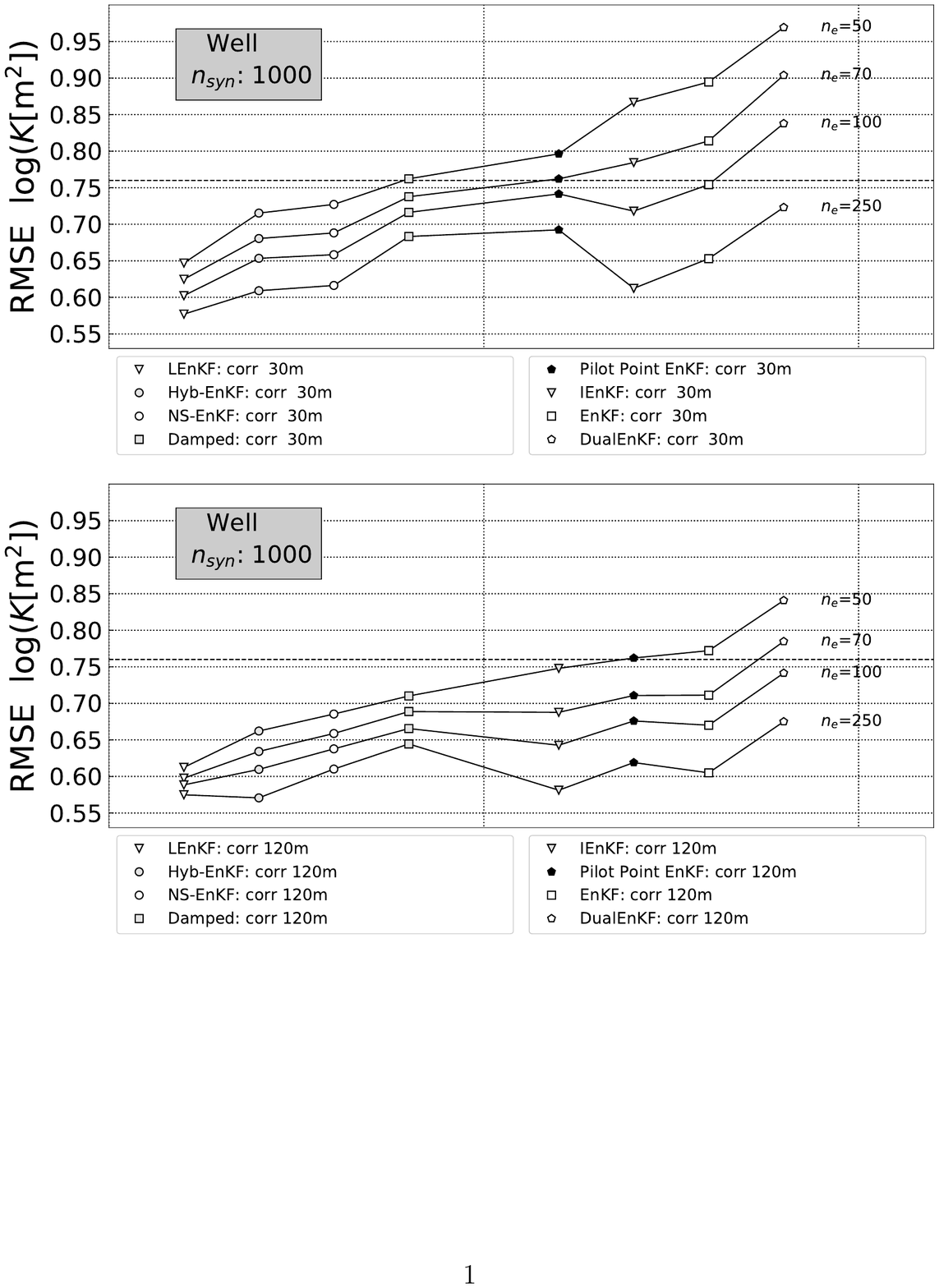}
    \caption{Comparison of the pilot point EnKF with other
      EnKF variants for the well model and correlation lengths
      $30\,\mathrm{m}$ (top) and $120\,\mathrm{m}$ (bottom). %
    }
  \label{fig:well-model-corr-comp}
\end{figure}%

Turning to the well model, the PP-EnKF for correlation length
$30\,\mathrm{m}$ is tested. %
Figure \ref{fig:well-model-corr-comp} (top) shows RMSEs calculated
over 1,000 synthetic experiments. %
Again, for all ensemble sizes, all EnKF methods yield slightly larger
RMSEs than for the case of the correct correlation length
($60\,\mathrm{m}$). %
The PP-EnKF ranks fifth among the EnKF methods for ensemble sizes 50
and 70. %
For ensemble size 70 it ranks sixth and for ensemble size 250
seventh. %
This almost reproduces the results for the correct correlation length,
only for ensemble size 250 the rank is slightly worse. %

We now look at results for the well model and correlation length
$120\,\mathrm{m}$ in Figure \ref{fig:well-model-corr-comp} (bottom). %
Here, for all ensemble sizes, all EnKF methods yield slightly smaller
RMSEs than for the standard case. %
Regarding the PP-EnKF, it ranks sixth for ensemble sizes 50, 70 and
250, and it ranks seventh for ensemble size 100. %
Thus, again the ranking of the PP-EnKF among the methods is slightly
worse than for the correct correlation length. %

In summary, the PP-EnKF is not affected more by a mis-specification of
the correlation length than other EnKF variants for both model setups,
even though the correlation length has a more pronounced role in the
PP-EnKF than in other variants. %
It has to be noted that the estimates the PP-EnKF may suffer from
inaccuracy caused by inaccurate prior covariance matrices. %
In the synthetic experiments of this study, we estimate the effect on
the PP-EnKF by deliberately choosing a slightly wrong initial mean of
permeability fields, as well as, in this section, wrong correlation
lengths in the construction of the correlation fields. %
Results for the PP-EnKF in these synthetic experiments are promising
and do not show a worsening compared to other EnKF variants. %
However, especially in realistic, large-scale model setups, the issue
of mis-specified correlations can arise for all EnKF variants and
possibly with additional severity. %

\subsection{Average rankings of the PP-EnKF}
\label{sec:overall-ranking}

In the last sections, various types of synthetic experiments were used
to compare the performance of the PP-EnKF to other EnKF methods. %
This included synthetic experiments with the correct and erroneous
prior correlation lengths, and repetitions for ensemble sizes of 50,
70, 100 and 250. %
In Table \ref{tab:average-rankings-rmse}, the average RMSE rankings
are displayed that are calculated from the twelve single RMSE rankings
(four ensemble sizes for each of three correlation lengths). %
For the tracer setup, the PP-EnKF has the third best average
ranking. %
Thus, for the tracer setup, the PP-EnKF has very good average results
outperformed only by the hybrid EnKF and the iterative EnKF. %
For this setup, the performance of the PP-EnKF justifies its usage
over the majority of other EnKF methods. %
For the well setup, the PP-EnKF only has the sixth best average
ranking. %
For this setup, looking at the average RMSE would not justify using
the PP-EnKF compared to existing EnKF variants. %

\begin{table}[h]
  \centering
  \caption{Average RMSE rankings of the eight EnKF methods for tracer
    and well setups. The average ranking is the mean of the twelve
    rankings from the RMSE comparisons of all tested prior correlation
    lengths.} %
  \label{tab:average-rankings-rmse}
  \begin{tabular}{l l l}
    \hline
    \qquad \qquad RMSE \qquad \qquad
    &
      Tracer \qquad \qquad
    &
      Well
    \\
    \hline
    EnKF
    &
      6.1667
    &
      6.3333
    \\
    Damped
    &
      5.9167
    &
      4.9167
    \\
    NS-EnKF
    &
      3.9167
    &
      3.3333
    \\
    DualEnKF
    &
      8.0
    &
      8.0
    \\
    Hyb-EnKF
    &
      1.0
    &
      1.9167
    \\
    LEnKF
    &
      5.4167
    &
      1.0833
    \\
    IEnKF
    &
      2.3333
    &
      4.5833
    \\
    \textbf{PP-EnKF}
    &
      3.25
    &
      5.833
    \\
    \hline
  \end{tabular}
\end{table} %

\begin{table}[h!]
  \centering
  \caption{Average STD rankings of the eight EnKF methods for the
    tracer and well setups. The average ranking is the mean of the
    four rankings from the overall STD comparison.} %
  \label{tab:average-rankings-std}
  \begin{tabular}{l l l}
    \hline
    \qquad \qquad STD \qquad \qquad
    &
      Tracer \qquad \qquad
    &
      Well
    \\
    \hline
    EnKF
    &
      4.75
    &
      8.0
    \\
    Damped
    &
      2.25
    &
      2.0
    \\
    NS-EnKF
    &
      7.75
    &
      2.5
    \\
    DualEnKF
    &
      6.0
    &
      6.0
    \\
    Hyb-EnKF
    &
      1.5
    &
      5.0
    \\
    LEnKF
    &
      4.25
    &
      3.5
    \\
    IEnKF
    &
      6.5
    &
      7.0
    \\
    \textbf{PP-EnKF}
    &
      3.0
    &
      2.0
    \\
    \hline
  \end{tabular}
\end{table} %

The reproduction of the correct ensemble variance is another key
performance for comparing the EnKF methods. %
Therefore, overall STD results for the four ensemble sizes 50, 70,
100, 250 were ranked and subsequently averaged, and results are
summarized in Table \ref{tab:average-rankings-std}. %
The PP-EnKF performs well among the eight tested methods. %
For the tracer setup, the PP-EnKF ranks third out of eight methods. %
In particular, the PP-EnKF ranks significantly better than the
iterative EnKF that ranked better for average RMSE. %
Thus, taking into account both, average RMSE and average STD, the
PP-EnKF is outperformed only by the hybrid EnKF. %
For the well setup, the PP-EnKF ranks first together with damped
EnKF. %
For the well setup and for small ensemble sizes, the PP-EnKF ranks
undivided first. %

\subsection{Variation of pilot point grids}
\label{sec:var-pp-grids}

One important degree of freedom added by the PP-EnKF is the choice of
the pilot points. %
In this section, RMSEs for different pilot point grids are compared in
the tracer and well model. %

\begin{figure}
  \centering
  \includegraphics[trim = 50 190 50 150, clip,
    width=\columnwidth]{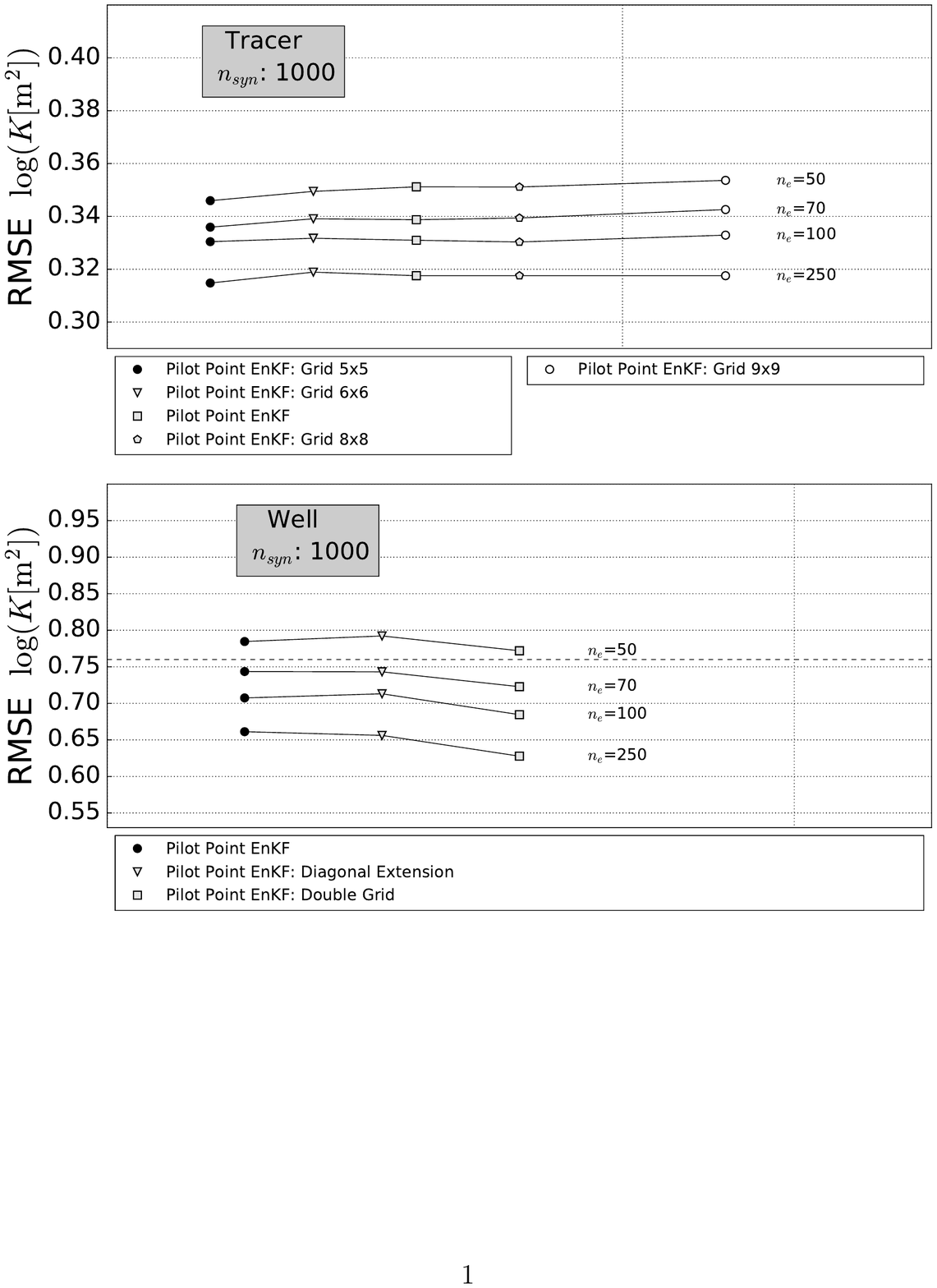}
  \caption{ Mean RMSE (calculated over $1,000$ synthetic experiments)
    for PP-EnKF with different regular grids for the tracer setup
    (top) and well setup (bottom). %
  }
  \label{fig:model-pp-grids}
\end{figure}%

For the tracer model, the compared grids are regular square grids with
different numbers of pilot points. %
All grids also include the two measurement locations of the tracer
model. %
The results in Figure \ref{fig:model-pp-grids} (top) show small
changes of the RMSE for different grids, all changes are smaller than
$0.01\, \log_{10}(K[\mathrm{m}^{2}])$. %

For the well model, the compared grids are extensions of the standard
grid, since the standard grid is made up of all measurement
locations. %
The first extension is adding pilot points on the diagonal between the
original pilot points. %
The second extension is another regular grid with double the number of
pilot points in each column and row. %
Again, the results in Figure \ref{fig:model-pp-grids} (bottom) exhibit
small changes of the RMSE, only the doubled grid yields significantly
smaller RMSE than the standard method. %

The results of the grid variation suggest two implications. %
First, the differences between the grids are not very large, thus at
least in the two setups treated here, the PP-EnKF shows a certain
robustness against different choices of regular grids in the
investigated range. %
Second, the results do also point in the direction that a smart choice
of grid might make a bigger difference in other setups. %
While the grid with the smallest number of pilot points yields the
smallest RMSEs for the relatively homogeneous synthetic true
permeability field of the tracer model, the grid with the largest
number of pilot points yields the smallest RMSEs for the more
heterogeneous well model. %
This suggests a model-specific number of pilot points required to
sufficiently approximate the real cross correlation functions. %
As a generic rule, we refer to the recommendations for pilot point
spacing in the literature that suggest about three pilot points per
range as a robust choice \citep{Capilla1997}. %
The range from \citet{Capilla1997} would correspond to two correlation
lengths in our study, a distance at which parameters are barely
correlated. %

In this study, we discussed the most straightforward placing of pilot
points on a regular grid. %
This pilot point placing is well adapted to this heterogeneous model
setup \citep{Hendricks2001, Doherty2010}. %
Additionally, there are ways to optimize the placing of pilot points
in the framework of optimal experimental design \citep{Mehne2011} or
using prior information \citep{Alcolea2006}. %
This could be adapted to the PP-EnKF by optimally placing pilot
points. %
Options include measures derived from the sample covariance, and the
pilot point locations could even be dynamically adapted during the
assimilation, for example after each updating step. %
In future research, it would be interesting to check the influence of
sophisticated pilot point placing methods on the performance of the
PP-EnKF, especially in larger, more realistic model setups than
discussed in this work. %
In summary, strategies to optimize the placing of pilot points are
documented in the literature. %
However, this was beyond the scope of this work and requires
substantial additional compute time. %
We followed here standard rules for placing of pilot points which gave
the documented satisfactory results and were not subject to large
changes in case of modifying the density of the pilot points. %
However, by optimizing the placing of the pilot points further
performance gain might be achieved. %

\subsection{Spurious correlation reduction}
\label{sec:spurious-correlation-reduction}

\begin{figure}
  \centering
  \includegraphics[trim = 50 270 70 240, clip,
    width=\columnwidth]{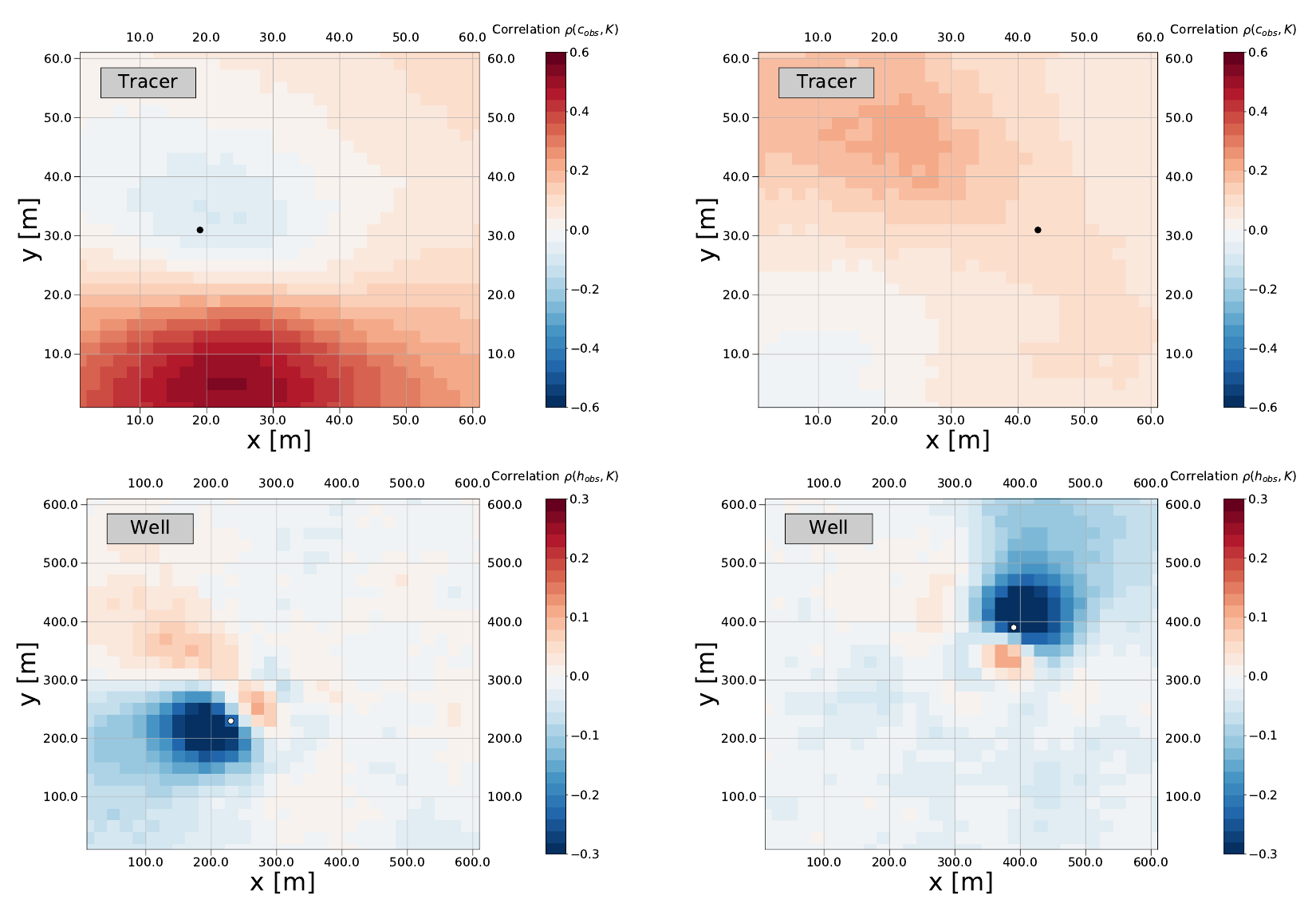}
    \caption{Reference correlations between an observed variable at a
      chosen observation location and the permeability throughout the
      model domain. %
      For the tracer setup (top), the two observation locations for
      concentration are plotted in black, for the well setup (bottom)
      the two (out of 49) observation points for head are plotted in
      white. %
    }
  \label{fig:correlation-fields}
\end{figure}%

Now the ability of the PP-EnKF to reduce spurious correlations is
tested by checking the correlations between observed variables and
updated permeabilities in the tracer and well setup after
assimilation. %
As a benchmark, we use a reference synthetic experiment with the EnKF
and 10,000 ensemble members. %
Figure \ref{fig:correlation-fields} shows the reference correlation
fields for the tracer and well setup after the EnKF updates. %
For the tracer setup, both observation locations are shown. %
The concentrations at the observation location in the highly permeable
region (observation location on the right, compare the synthetic true
permeability fields from Figure \ref{fig:true-perm}) yields comparably
small correlations. %
The observation location in the low-permeable region (observation
location on the left) yields large positive correlation with the
low-permeability region south of the observation location. %
For the EnKF update, this positive correlation implies that larger
measured tracer concentrations will lead to updates towards larger
permeability values (promoting tracer transport towards the
measurement location). %
Turning to the well setup, two representative head observation
locations are displayed. %
One can observe a general trend, small positive correlations in the
direction of the injection well in the center of the model and large
negative correlations in the direction of the model boundaries. %
For the EnKF update, this means that higher measured hydraulic heads
(smaller drawdowns) will lead to larger permeabilities in the center
of the model (promoting flow towards the measurement location) and
lower permeabilities further away from the center (prohibiting flow
away from the measurement location). %
Note that the correlation structure in the well setup is finer and
more pronounced than in the tracer setup. %
This is related partly to the characteristics of the permeability
fields. %

\begin{table}[h]
  \centering
  \caption{Root-mean-square errors of correlation fields (compared to
    reference correlation fields).} %
  \label{tab:correlation-rmses}
  \begin{tabular}{l c c c c c c}
    \hline
    &
      RMSEs
    &
    &
    &
    &
    &
      Mean
    \\
    \hline                      %
    Tracer (EnKF)
    &
      0.133
    &
      0.167
    &
      0.215
    &
      0.139
    &
      0.154
    &
      0.161
    \\
    &
      0.163
    &
      0.266
    &
      0.106
    &
      0.125
    &
      0.141
    &
    \\
    Tracer (PP-EnKF)
    &
      0.142
    &
      0.169
    &
      0.213
    &
      0.128
    &
      0.155
    &
      0.162
    \\
    &
      0.161
    &
      0.260
    &
      0.113
    &
      0.137
    &
      0.137
    &
    \\
    \hline                      %
    Well (EnKF)
    &
      0.180
    &
      0.159
    &
      0.168
    &
      0.172
    &
      0.178
    &
      0.173
    \\
    &
      0.173
    &
      0.171
    &
      0.190
    &
      0.162
    &
      0.173
    &
    \\
    Well (PP-EnKF)
    &
      0.152
    &
      0.153
    &
      0.145
    &
      0.149
    &
      0.158
    &
      0.149
    \\
    &
      0.142
    &
      0.149
    &
      0.147
    &
      0.145
    &
      0.147
    &
    \\
    \hline
  \end{tabular}
\end{table} %

Now we compare correlation fields from 10 synthetic experiments (for
the PP-EnKF and the classical EnKF using ensemble size 50) to the
correlation fields of the reference. %
Table \ref{tab:correlation-rmses} shows the RMSE values for each
synthetic experiment. %
For the tracer setup, for five synthetic experiments, PP-EnKF is
closer to the reference than for EnKF, and for five synthetic
experiments the opposite holds. %
The mean RMSEs for the PP-EnKF and the EnKF for the ten synthetic
experiments are very close. %
For the well setup, PP-EnKF has smaller RMSE-values than the EnKF for
all ten synthetic cases, with on average a smaller RMSE of 13.9\%. %

The results from this section support results from the overall
standard deviation STD from Figure \ref{fig:mean-std-comps}. %
For the tracer setup, the similar results from this section correspond
to relatively similar overall STDs of the EnKF and the PP-EnKF in
Figure \ref{fig:mean-std-comps}. %
For the well setup, the PP-EnKF resulted in a larger STD compared to
the EnKF. %
In this section, we additionally find a better characterization of the
spatial correlation structure by PP-EnKF, compared to EnKF. %
Thus, the larger overall uncertainty shown by the STD does not stem
from additional noise, it is rather the result of a correlation
structure closer to a large-ensemble run. %
To summarize results for the synthetic experiments for ensemble size
50, the PP-EnKF has both a better uncertainty characterization and a
better RMSE than the EnKF. %
This illustrates that the PP-EnKF shows significant advantages
compared to the EnKF. %

\section{Conclusion}
\label{sec:conclusion}

The ensemble Kalman filter is a powerful tool for parameter estimation
used in the geosciences. %
In this work, we discuss a variant of the EnKF, the PP-EnKF that aims
to reduce spurious covariances for small ensemble sizes. %
The PP-EnKF was originally introduced and discussed by
\citet{Heidari2013} and \citet{Tavakoli2013}. %
In these publications, the performance of the PP-EnKF is evaluated
yielding some promising results. %
However, it remained unclear whether the method yields clear
advantages in terms of performance over other methods. %
Starting with a thorough mathematical exposition of the PP-EnKF, we
investigated the performance of the method regarding reproduction of
reference parameter fields, standard deviation of the ensemble, and
reduction of spurious correlation. %

The way the PP-EnKF stabilizes the covariance matrix can be compared
with the hybrid EnKF and the local EnKF. %
We claim that the main theoretical advantage of the PP-EnKF compared
to these methods is that its update at pilot points is computed from
the unmodified ensemble covariance matrix. %
In the local EnKF and the hybrid EnKF, the covariance matrix is
explicitly modified. %
The PP-EnKF introduces two additional input parameters compared to
other EnKF methods, the locations of the pilot points, and the
covariance matrix of the kriging interpolation. %
Pilot points on a regular subgrid of the model domain are
investigated, with little differences between the results for
different pilot point densities. %
All measurement locations are included as pilot point locations. %
In models, one could vary the density of pilot point locations
according to given prior information. %
In regions, where large updates are expected, the number of pilot
points can be increased. %
The kriging covariance matrix of the PP-EnKF is defined according to
the prior variogram model, which is also used as input to generate
random fields for all EnKF methods. %

The PP-EnKF compares well to other EnKF methods for two physical model
setups, a solute transport model and a model around an injection
well. %
This is concluded from an extensive comparison of RMSEs between
parameter estimation results from $1,000$ synthetic experiments and a
synthetic true parameter field. %
Even for synthetic experiments with erroneous prior correlation
lengths (half and twice the correct correlation length), the
performance of the PP-EnKF remains similar to the case with the
correct prior correlation length. %
This is important, as the prior correlation length plays a special
role in the PP-EnKF method. %
In the average RMSE-ranking, the PP-EnKF is third best for the tracer
and sixth best for the well setup. %
Compared to the EnKF, the PP-EnKF is performing significantly better
in both setups. %

The PP-EnKF ranks particularly well against the other EnKF methods
regarding the preservation of the spatial variability throughout the
estimation. %
Especially, for ensembles sizes of 50 and 70 and for the well model,
the PP-EnKF yields the ensemble variance that compares best to a test
run with the EnKF and a very large ensemble size of 10,000. %
In an average overall STD ranking, the PP-EnKF ranks third best for
the tracer setup and best in the well setup. %
Additionally, distributed correlation fields of the PP-EnKF and the
classical EnKF were compared. %
For the well setup, the correlations of the PP-EnKF are significantly
closer to a reference field than the correlations of the classical
EnKF. %
For the tracer setup, the correlations of the PP-EnKF and the
classical EnKF are equally close to the reference. %
Reproducing the posterior variance is an important feature of an
EnKF method, since many EnKF methods suffer from an underestimation of
the posterior variance that may lead to wrong interpretation of
results or in the worst case to filter divergence. %
The PP-EnKF not only ranks particularly well against the other
EnKF methods regarding the reproduction of the posterior variance. %
For a small ensemble, it also reproduces spatially distributed
correlation fields better than the classical EnKF. %
This suggests that the PP-EnKF is able to preserve better than the
EnKF not only the variance in the permeability field, but also the
correlations of the permeability field with the dynamic variables. %

From the aforementioned discussions we conclude that the PP-EnKF
outperforms clearly the standard EnKF, and outperforms most
EnKF variants for the reproduction of the ensemble spread. %
It is therefore a very interesting EnKF variant, with the need for
further research to investigate issues like the optimal placing of
pilot points and other applications like the estimation of soil
hydraulic parameters. %

\newpage
\section{Acknowledgements}
\label{sec:acknowledgements}
This study was supported by the Deutsche
Forschungsgemeinschaft. Simulations were performed with computing
resources granted by RWTH Aachen University under project rwth0009. %
The data used will be made available after acceptance of the
manuscript through a data repository (Platform: Zenodo). %


\end{document}